\documentclass{ws-procs9x6}
\usepackage{graphicx}
\usepackage{latexsym}

\newcommand{\includefigure}[2]{ {\centering
        \includegraphics*[scale=#1]{#2}}}

\newcommand{\mh}{\mbox{$M_H$}}   
\newcommand{\msb}{\ensuremath{\overline{\rm{MS}}}}       
\newcommand{\oh}{\ensuremath{\frac{1}{2}}}                                          
\newcommand{\vev}[1]{\ensuremath{\langle 0|#1|0\rangle}}
\newcommand{\lag}{\ensuremath{\mathcal{L}}}
\newcommand{\sthto}{\ensuremath{SU(3) \times SU(2) \times U(1)}}                                                                      
                      
\newcommand{\sto}{\ensuremath{SU(2) \times U(1)}}                                       
\newcommand{\st}{\ensuremath{SU(2)}}                                                                                       
                                                                                       
\newcommand{\eeql}[1]{\label{#1}\eeq}
\newcommand{\vp}{\ensuremath{\phi}}
\newcommand{\beq}{\begin{equation}}                                             
\newcommand{\eeq}{\end{equation}}     
\newcommand{\lagl}{Lagrangian}
\newcommand{\her}{Hermitian}

\newcommand{\ra}{\rightarrow}
\newcommand{\RA}{\rightarrow}

\newcommand{\p}[1]{\ensuremath{10^{#1}}}
\newcommand{\CP}{$CP$}
\newcommand{\refl}[1]{(\ref{#1})}

\begin{document}

\title{INTRODUCTION TO THE STANDARD MODEL AND ELECTROWEAK PHYSICS}

\author{PAUL LANGACKER}

\address{School of Natural Sciences,
Institute for Advanced Study \\
Princeton, NJ 08540, USA\\
E-mail: pgl@ias.edu}

\begin{abstract}
A concise introduction is given to the standard model, including 
the structure of the
QCD and electroweak Lagrangians, spontaneous symmetry breaking,
experimental tests, and problems.
\end{abstract}


\bodymatter

\section{The Standard Model Lagrangian}

\subsection{QCD}
 
The standard model (SM) is a gauge theory~\cite{Weyl:1929fm,Yang:1954ek} of the microscopic
interactions. The strong interaction part, quantum chromodynamics (QCD)\footnote{See~\refcite{Gross:2005kv} for a historical overview. Some recent reviews
include~\refcite{Bethke:2006ac} and the QCD review in~\refcite{Amsler:2008zz}.} is 
an $SU(3)$ gauge theory described by the
Lagrangian density \beq \lag_{SU(3)} = - \frac{1}{4} F^i_{\mu\nu} F^{i\mu\nu} + \sum_r
\bar{q}_{r\alpha} i \not{\!\!D}^\alpha_\beta \,q^\beta_r, \label{eq1:c20}
\eeq
where $g_s$ is the QCD gauge coupling
constant,
\beq F^i_{\mu\nu} = \partial_\mu G^i_\nu - \partial_\nu G^i_\mu -
g_s f_{ijk}\; G_\mu^j\; G_\nu^k  \label{eq1:c21} \eeq
is the field strength tensor for the gluon
fields $G^i_\mu, \; i = 1, \cdots, 8$,  and the structure constants $f_{ijk}$
 $ (i, j, k = 1, \cdots, 8)$ are
defined by
\beq [\lambda^i, \lambda^j] = 2 i f_{ijk} \lambda^k, \eeq
where the $SU(3)$ $\lambda$ matrices are defined in
Table~\ref{table1:000b}. The $\lambda$'s are normalized by
$\text{Tr}\, \lambda^i \lambda^j = 2 \delta^{ij}$, so that  
$\text{Tr}\,  [\lambda^i, \lambda^j] \lambda^k=4i f_{ijk}$.

\begin{figure}
\centering
\includefigure{0.6}{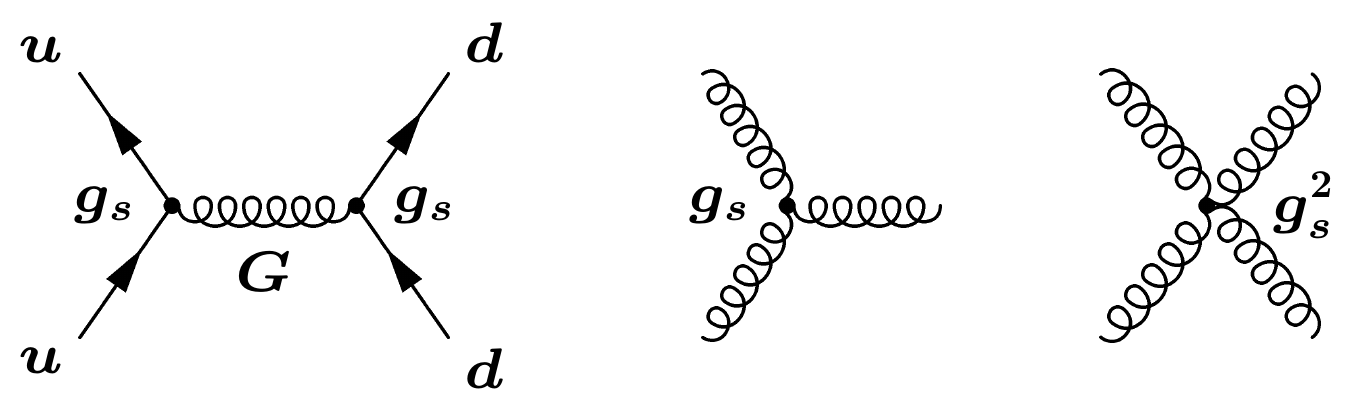}
\caption{Interactions in QCD.}
\label{fe5_qcd}
\end{figure}

    The $F^2$ term leads to three and four-point gluon
self-interactions, shown schematically in Figure~\ref{fe5_qcd}. 
The second term in $\lag_{SU(3)}$ is the gauge covariant derivative for the
quarks: $q_r$ is the $r^{  th}$ quark flavor, $\alpha, \beta = 1,2,3$ are
color indices, and \beq D^\alpha_{\mu \beta} = (D_\mu)_{\alpha \beta} =
\partial_\mu \delta_{\alpha \beta} + i g_s \;G^i_\mu\; L^i_{\alpha\beta},
\label{eq1:c22} \eeq
where the quarks transform according to the triplet representation matrices
$L^i ={\lambda^i}/{2}$.
The color interactions are diagonal in the flavor
indices, but in general change the quark colors. They are purely vector
(parity conserving). There are no bare mass terms for the quarks in
(\ref{eq1:c20}). These would be allowed by QCD alone, 
but are forbidden by the chiral
symmetry of the electroweak part of the theory.  The quark masses will be
generated later by spontaneous symmetry breaking.  There are in addition
effective ghost and gauge-fixing terms
which enter into the quantization of both the $SU(3)$ and
electroweak \lagl s, and there is the possibility of adding an (unwanted) term
which violates $CP$ invariance.

\begin{table}
\tbl{The $SU(3)$ matrices.}
{\begin{tabular}{rrr} \toprule
$\lambda^i = \left( \begin{array}{cc} \tau^i & 0 \\ 0 & 0
\end{array} \right),$ & $i = 1,2,3$ &\\
$\lambda^4 = \left( \begin{array}{ccc} 0 & 0 & 1 \\ 0 & 0 & 0 \\
1 & 0 & 0 \end{array} \right)$ & &
\ \ \ $\lambda^5 = \left( \begin{array}{lcr} 0 & 0 & -i \\ 0 & 0
& 0 \\ i & 0 & 0 \end{array} \right)$ \\
$\lambda^6 = \left( \begin{array}{ccc} 0 & 0 & 0 \\ 0 & 0 & 1 \\
0 & 1 & 0 \end{array} \right)$ & &
$\lambda^7 = \left( \begin{array}{lcr} 0 & 0 & 0 \\ 0 & 0 & -i
\\ 0 & i & 0 \end{array} \right)$ \\
$\lambda^8 = \frac{1}{\sqrt{3}} \left( \begin{array}{lcr} 1 & 0 &
0 \\ 0 & 1 & 0 \\ 0 & 0 & -2 \end{array} \right)$ & & \\ \botrule
\end{tabular}}
\label{table1:000b}
\end{table}

\begin{figure}[htbp]
\begin{minipage}[t]{5.0cm} 
\includefigure{0.34}{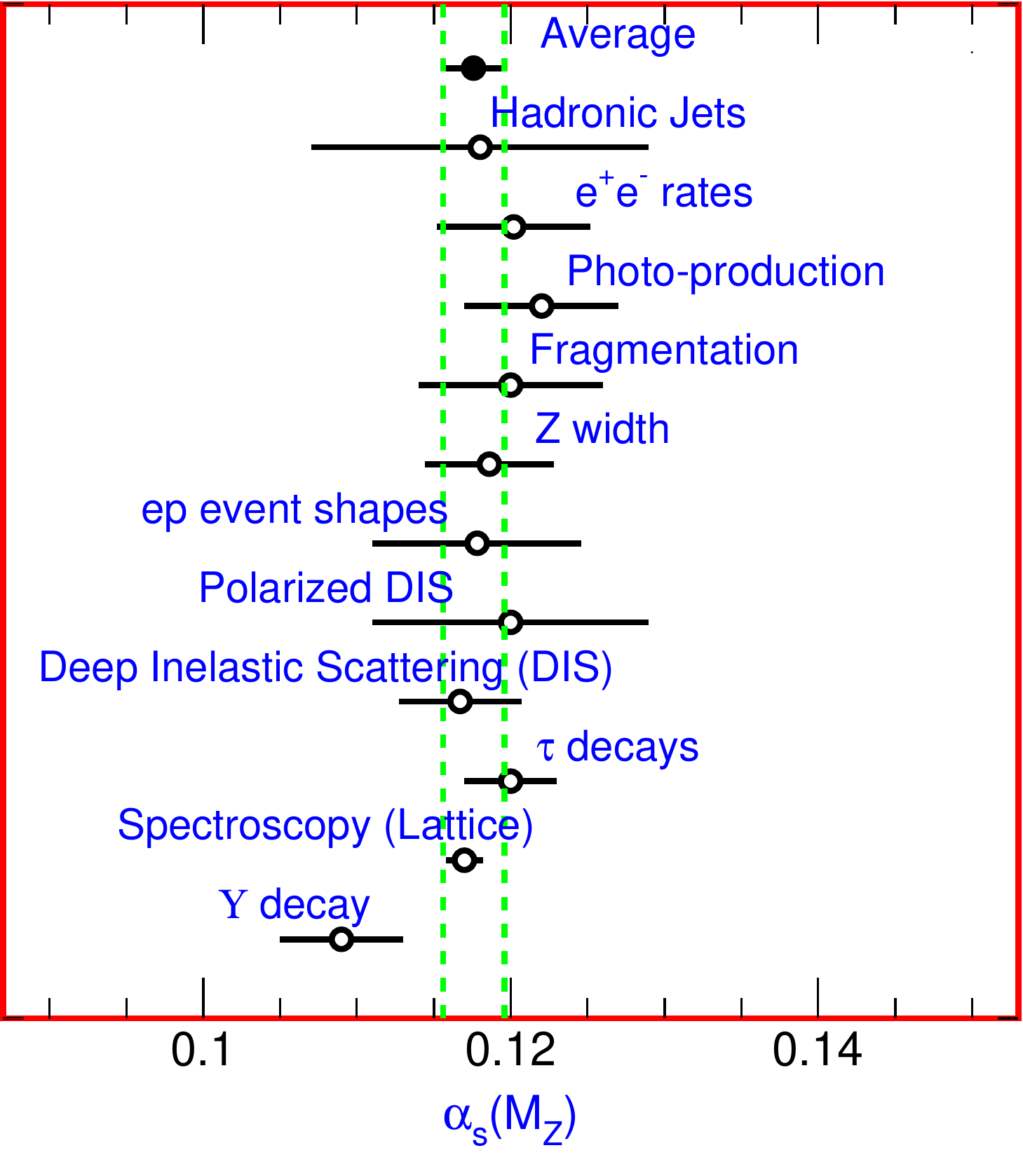}
\end{minipage}
\hspace*{0.2cm}
\begin{minipage}[t]{5.0cm} 
\includefigure{0.34}{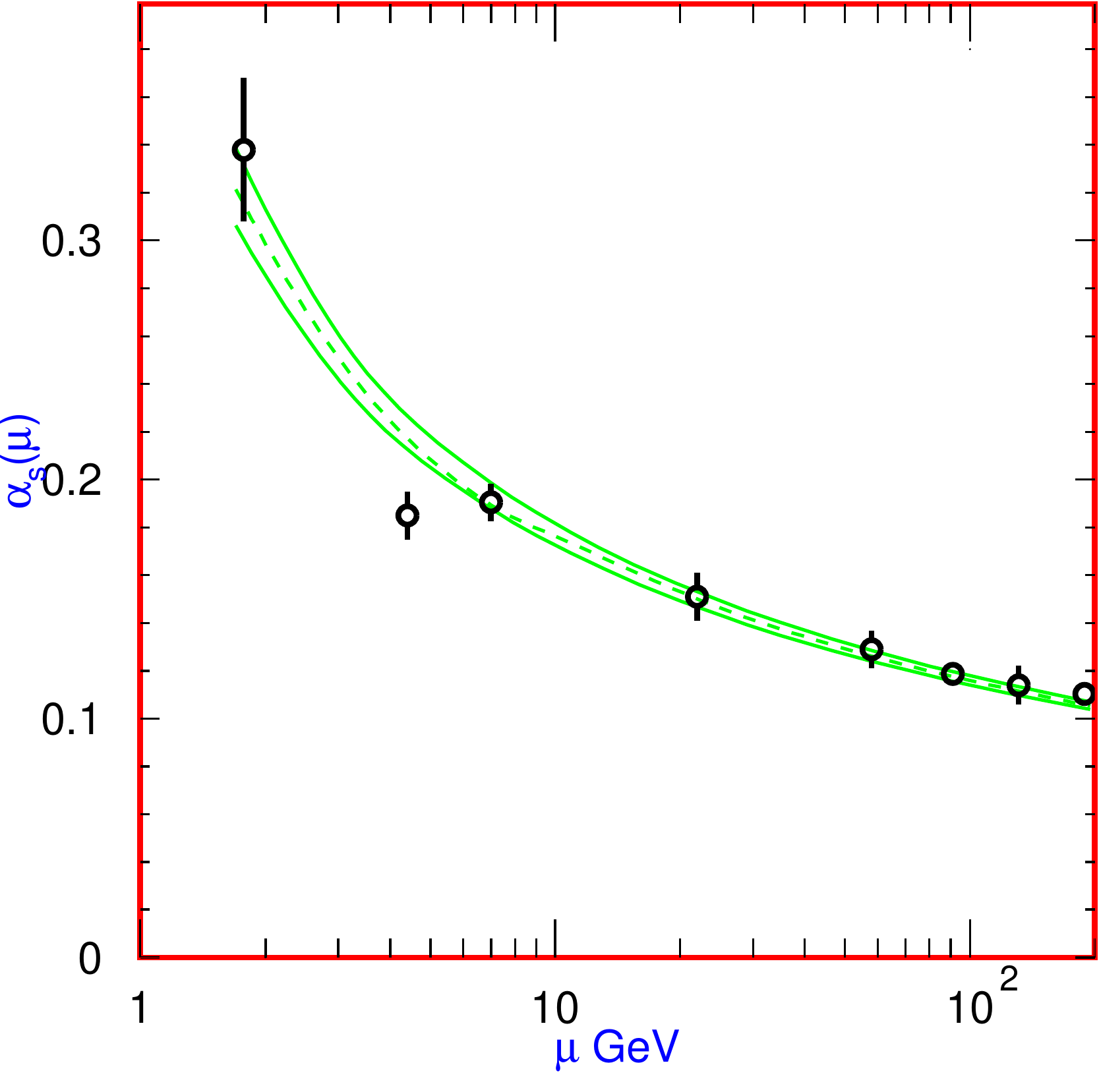}
\end{minipage}
\caption{Running of the QCD coupling $\alpha_s(\mu)=g_s(\mu)^2/4\pi$. Left: various experimental determinations extrapolated to $\mu=M_Z$ using QCD. Right:  experimental values
plotted at the $\mu$ at which they are measured. The band is the best fit
QCD prediction.
Plot courtesy of the Particle Data Group~\cite{Amsler:2008zz}, {\tt http://pdg.lbl.gov/}.}
\label{qcdtests}
\end{figure}

QCD has the property of asymptotic freedom~\cite{Gross:1973id,Politzer:1973fx}, i.e., the running coupling becomes
weak at high energies or short distances. It has been extensively
tested in this regime, as is illustrated in Figure~\ref{qcdtests}. At low energies or large distances it becomes strongly
coupled (infrared slavery)~\cite{Fritzsch:1973pi}, presumably leading to the confinement of 
quarks and gluons. QCD incorporates the observed global symmetries
of the strong interactions, especially the spontaneously broken global $SU(3) \times SU(3)$
 (see, e.g., \refcite{Gasser:1982ap}).

\subsection{The Electroweak Theory}
The electroweak theory~\cite{Glashow:1961tr,Weinberg:1967tq,Salam:1968rm} is based on the $SU(2) \times U(1)$ \lagl\footnote{For a recent discussion, see the electroweak  review in~\refcite{Amsler:2008zz}.}
\beq \lag_{SU(2) \times U(1)} = \lag_{gauge} + \lag_\phi+   \lag_f  + 
\lag_{Yuk}. \label{eqch10b} \eeq
The gauge part is
\beq \lag_{  gauge} = - \frac{1}{4} W^i_{\mu \nu} W^{\mu \nu i} -
\frac{1}{4} B_{\mu \nu} B^{\mu \nu}, \label{eqch11b} \eeq
where $W^i_\mu, \; i = 1, \; 2, \; 3$ and
$B_\mu$  are respectively the $SU(2)$ and $U(1)$ gauge fields,
with field  strength tensors 
\begin{eqnarray} B_{\mu \nu} &=& \partial_\mu B_\nu - \partial_\nu
B_\mu \nonumber \\
W^i_{\mu \nu} &=& \partial_\mu W_\nu^i - \partial_\nu W_\mu^i - g
\epsilon_{ijk} W^j_\mu W^k_\nu, \label{eqch12a} \end{eqnarray}
where $g (g')$ is the $SU(2)$ $(U(1))$ gauge coupling and $\epsilon_{ijk}$ is
the totally antisymmetric symbol.
 The $SU(2)$ fields
have three and four-point self-interactions.
$B$ is a $U(1)$ field associated with the
weak hypercharge 
$ Y = Q - T^3$, where $Q$ and $T^3$ are respectively the electric charge
operator and the third component of weak $SU(2)$.
(Their eigenvalues  will be denoted by $y$, $q$, and $t^3$, respectively.) It has no
self-interactions. The $B$ and $W_3$ fields will eventually mix to form the
photon and $Z$~boson.
 
The scalar part of the \lagl\ is
\beq \lag_\phi  = (D^\mu \phi )^{\dag} D_\mu \phi  - V(\phi ),
\label{eqch13b} \eeq
where $\phi  = \left(\begin{array}{c} \phi ^+ \\ \phi ^0 \end{array}
\right)$ is a complex Higgs scalar, which is a doublet under $SU(2)$ with $U(1)$
charge $y_\phi  = + \frac{1}{2} $.  The gauge covariant derivative is
\beq D_\mu \phi  = \left( \partial_\mu + i g \frac{\tau^i}{2}
W_\mu^i + \frac{i g'}{2} B_\mu \right) \phi , \label{eqch14b} \eeq
where the $\tau^i$ are the Pauli matrices.
   The square of the covariant derivative leads to three and four-point
interactions between the gauge and scalar fields.
 
$V(\phi )$ is the Higgs potential.  The combination of $SU(2) \times U(1)$
invariance and renormalizability restricts $V$ to the form
\beq V(\phi ) = + \mu^2 \phi ^{\dag}\phi  + \lambda (\phi ^{\dag}
\phi )^2. \label{eq15} \eeq
For $\mu^2 < 0$ there will be spontaneous symmetry breaking.  The
$\lambda$ term describes a quartic  self-interaction
between the scalar fields. Vacuum stability requires $\lambda > 0$.
 
The fermion term is
\beq \begin{split}
\lag_f =& \sum^F_{m = 1} \left( \bar{q}^0_{mL} i
\not\!\!D q^0_{mL} + \bar{l}^0_{mL} i \not\!\!D l^0_{mL}
 + \bar{u}^0_{mR} i \not\!\!D u^0_{mR} \right.\\ 
 & \quad +\left.
\bar{d}^0_{mR} i \not\!\!D d^0_{mR} + \bar{e}^0_{mR} i \not\!\!D
e^0_{mR}+ \bar{\nu}^0_{mR} i  \not\!\!D
\nu^0_{mR} \right).
\end{split} \label{eqch16} \eeq
In (\ref{eqch16}) $m$ is the family index, $F \ge 3$ is the number
of families, and $L(R)$ refer to the left (right) chiral projections
$\psi_{L(R)} \equiv (1 \mp \gamma_5) \psi/2$. The left-handed quarks
and leptons 
\beq
q^0_{mL}= \left( \begin{array}{c} u^0_m \\ d^0_m \end{array}
\right)_L \ \ \ \ \ l^0_{mL} = \left( \begin{array}{c} \nu^0_m \\ e^{-0}_m
\end{array} \right)_L
\eeq
transform as $SU(2)$ doublets, while the right-handed fields
$u^0_{mR}, \; d^0_{mR}$,  $e^{-0}_{mR}$, and $\nu^0_{mR}$ are singlets. 
Their $U(1)$ charges are  $y_{q_L} = \frac{1}{6}, \; y_{l_L} =-\frac{1}{2}, \;
y_{\psi_R} = q_\psi$. The superscript $0$ refers to the weak 
eigenstates, i.e., fields
transforming according to definite $SU(2)$ representations.  They may
be mixtures of mass eigenstates (flavors).
The quark color indices $\alpha = r,
\; g, \; b$ have been suppressed. 
The gauge covariant derivatives are
\begin{equation} 
\begin{array}{lclclcl}
D_\mu q^0_{mL} &=& \left( \partial_\mu + \frac{i g}{2}\vec \tau\cdot \vec W_\mu 
+ \frac{ ig'}{6} B_\mu \right) q^0_{mL}
&  \ \ \ \ \ &
D_\mu u^0_{mR} &=& \left( \partial_\mu + \frac{2ig'}{3} B_\mu \right)
u^0_{mR}\vspace*{2pt}  \\
D_\mu l^0_{mL} &=& \left( \partial_\mu + \frac{i g}{2}\vec \tau\cdot \vec W_\mu 
- \frac{i g'}{2} B_\mu \right) l^0_{mL}
&  \ \ \ \ \ &
D_\mu d^0_{mR} &=& \left(\partial_\mu -\frac{ig'}{3}B_\mu  \right)
d^0_{mR} \vspace*{2pt} \\
& &  &  \ \ \ \ \ &
D_\mu e^0_{mR} &=& \left( \partial_\mu - i g' B_\mu     \right)
e^0_{mR}\vspace*{2pt}\\
& &  &  \ \ \ \ \ & 
D_\mu \nu^0_{mR} &=& \partial_\mu     \nu^0_{mR},
\end{array} 
 \label{eqch17} \end{equation}
from which one can read off the gauge interactions between the $W$ and $B$
and the fermion fields.
 The different transformations of the $L$ and $R$ fields (i.e.,  the symmetry
is chiral) is the origin of parity violation in the electroweak sector. 
The chiral symmetry also forbids
any bare mass terms for the fermions.
We have tentatively included \st-singlet right-handed neutrinos $ \nu^0_{mR} $ in \refl{eqch16}, because they
are required in many models for neutrino mass. However, they are not necessary for the consistency of the
theory or for some models of neutrino mass, and it is not certain whether they exist or are part of the low-energy theory.

The standard model is anomaly free  for the assumed fermion content.
There are no $SU(3)^3$ anomalies because the quark assignment is non-chiral, and no $\st^3$ anomalies because the representations are real. 
The $SU(2)^2Y$ and $Y^3$
anomalies cancel between the quarks and leptons in each family, by what appears to be an accident. The  $SU(3)^2Y$ and 
$Y$  anomalies cancel between the $L$ and $R$ fields, ultimately because the hypercharge
assignments are made in such a way that $U(1)_Q$ will be non-chiral.

The last term in (\ref{eqch10b}) is 
\beq
\begin{split}
 \lag_{Yuk} &= - \sum^F_{m,n =1} \left[
\Gamma^u_{mn} \bar{q}\,^0_{mL} \tilde{\vp} u^0_{nR} + \Gamma^d_{mn}
\bar{q}\,^0_{mL} \vp d^0_{nR} \right. \\
&+ \left. \Gamma^e_{mn} \bar{l}\,^0_{mn} \vp e^0_{nR}\ +
\Gamma^\nu_{mn} \bar{l}\,^0_{mL} \tilde{\vp} \nu^0_{nR} \right] + h.c.,
\end{split}
 \label{eqch18} \eeq
where the matrices $\Gamma_{mn}$  describe the  Yukawa
couplings between the single Higgs doublet, $\phi $, and the various
flavors $m$ and $n$ of quarks and leptons.  One needs representations of
Higgs fields with
 $y = +\frac{1}{2}$ and $-\frac{1}{2}$ to
give masses to the down quarks and electrons ($+\frac{1}{2}$), and to the up
quarks and neutrinos  ($-\frac{1}{2}$).  The representation $\phi ^{\dag}$ has $y = - \frac{1}{2}$,
but transforms as the $2^*$ rather than the 2.  However, in
                   $SU(2) $              the $2^*$
representation is related to the 2 by a similarity
transformation, and
$\tilde{\phi } \equiv i \tau^2 \phi ^{\dag} = \left( \begin{array}{c}
\phi ^{0^{\dag}} \\ - \phi ^- \end{array} \right)$ transforms as a 2
with $y_{\tilde{\phi }}=-\frac{1}{2}$. All
of the masses can therefore be generated with a single Higgs doublet if
one makes use of both $\phi $ and $\tilde{\phi }$.  The fact that the
fundamental and its conjugate are equivalent does not generalize to higher
unitary groups.  Furthermore, in supersymmetric extensions of the standard
model  the supersymmetry forbids the use of a single Higgs doublet in both
ways in the \lagl, and one must add a second Higgs
doublet. Similar statements apply to most theories with an
additional $U(1)'$ gauge factor, i.e., a heavy $Z'$ boson.

\section{Spontaneous Symmetry Breaking}

Gauge invariance (and therefore renormalizability) does not allow
mass terms in the \lagl \ for the gauge bosons or for chiral fermions.
Massless gauge bosons are not acceptable for the weak interactions, 
which are known to be short-ranged. Hence, the gauge invariance must be
broken spontaneously~\cite{Schwinger:1962tn,Anderson:1963pc,Higgs:1964ia,Higgs:1966ev,Englert:1964et,Guralnik:1964eu}, which preserves
the renormalizability~\cite{Hooft:1971rn,Hooft:1972ue,Lee:1972fj,Lee:1972fn}. The idea is that the lowest energy
(vacuum) state does not respect the gauge symmetry and induces
effective masses for particles propagating through it.
 
Let us introduce the
complex vector
\beq v = \langle 0 | \phi  | 0 \rangle = \text{ constant}, \label{eqch20} \eeq
which has components that are  the vacuum
expectation values of the various complex scalar fields.
$v$ is determined  by rewriting the Higgs potential
      as a function of $v$, $V(\phi ) \ra V(v)$, and choosing $v$
such that $V$ is minimized.  That is, we
interpret $v$   as the lowest energy
solution  of the classical  equation of motion\footnote{It suffices to
consider constant $v$ because
     any space or time dependence $\partial_\mu v$
would increase the energy of the solution. Also, one can take
$ \langle 0 | A_\mu | 0
\rangle = 0,$
because any non-zero vacuum value for a higher-spin field would
violate Lorentz invariance. However, these extensions are involved in higher energy classical solutions (topological defects),  such as monopoles, strings, domain
walls, and textures~\cite{Coleman:107318,Vilenkin:1984ib}.}.         
The quantum theory is
 obtained by
considering fluctuations around this classical minimum,  $\phi  = v +
\phi '$.
 
\begin{figure}
\centering
\includegraphics*[scale=0.8]{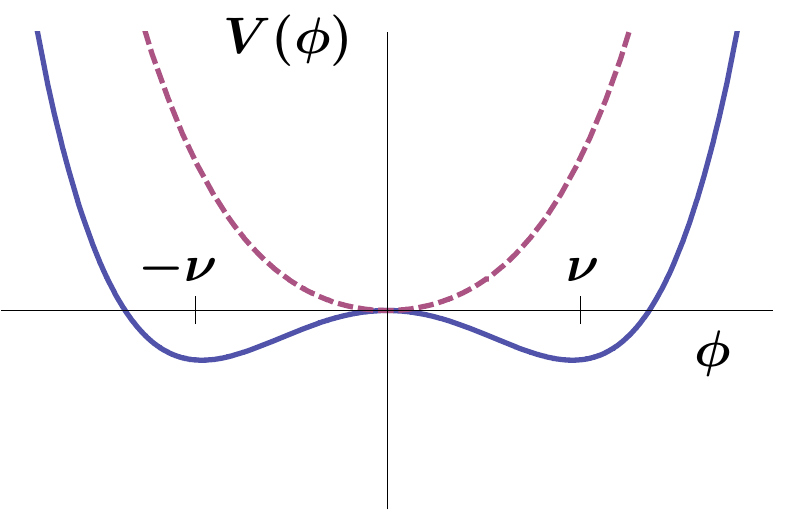}
\caption{The Higgs potential $V(\phi )$ for
$\mu^2 > 0$ (dashed line) and $\mu^2 < 0$ (solid line).}
\label{figure11}
\end{figure}

The single complex Higgs
doublet in  the standard model can be rewritten in a Hermitian basis as
\beq \phi  = \left( \begin{array}{c} \phi ^+ \\ \phi ^0
\end{array} \right) = \left( \begin{array}{c} \frac{1}{\sqrt{2}} (
\phi _1 - i \phi _2)        \\       \frac{1}{\sqrt{2}}(\phi _3- i
\phi _4  \end{array} \right), \label{eqch22} \eeq
where $\phi _i = \phi _i^{\dag}$
represent four \her\ fields.  In this new basis the Higgs potential
becomes
\beq V(\phi ) = \frac{1}{2} \mu^2 \left( \sum^4_{i=1} \phi ^2_i \right) +
\frac{1}{4} \lambda \left( \sum^4_{i=1} \phi ^2_i \right)^2
, \label{eqch23} \eeq
which is clearly $O(4)$ invariant.  Without loss
of generality we can choose the axis   in this four-dimensional space
so that $\langle 0| \phi _i |0 \rangle = 0, \;\ i = 1, 2, 4 $
and $\langle 0 | \phi _3 | 0 \rangle = \nu$. Thus,
\beq V (\phi ) \ra V(v) = \frac{1}{2} \mu^2 \nu^2 + \frac{1}{4}
\lambda \nu^4, \label{eqch24} \eeq
which must be minimized with respect to $\nu$.  Two important cases
are illustrated in Figure~\ref{figure11}.  For $\mu^2 > 0$ the minimum
occurs at $\nu = 0$.  That is, the vacuum is empty space and $SU(2) \times
U(1)$ is unbroken at the minimum.  On the other hand, for $\mu^2< 0$
the $\nu = 0$ symmetric point is unstable, and the minimum occurs at
some nonzero value of $\nu$ which breaks the $SU(2) \times U(1)$ symmetry.
The point is found by requiring
\beq V' (\nu) = \nu (\mu^2 + \lambda \nu^2) = 0,\label{eqch25} \eeq
which has the solution
$\nu = \left(      {-\mu^2}/{\lambda} \right)^{1/2} $
at the minimum.  (The solution for $-\nu$ can also be transformed into
this standard form by an appropriate $O(4)$ transformation.)
The dividing point $\mu^2 = 0$ cannot be treated classically.  It
is necessary to consider the one loop corrections to the potential, in
which case it is found that the symmetry is again spontaneously
broken~\cite{Coleman:1973jx}.
 
We are interested in the case $\mu^2 < 0$, for which the Higgs doublet
is replaced, in first approximation, by its classical value
$ \phi  \ra \frac{1}{\sqrt{2}} \left( \begin{array}{c} 0 \\ \nu
\end{array} \right) \equiv v$.
The generators $L^1$, $L^2$, and $L^3 - Y$ are spontaneously
broken ({e.g.,} $L^1 v \neq 0$).  On the other hand, the vacuum carries no
electric charge ($Qv = (L^3 + Y) v = 0$),
so the $U(1)_{Q}$ of electromagnetism is not broken.  Thus,
the electroweak $SU(2) \times U(1)$ group is spontaneously broken to the $U(1)_{Q}$ subgroup,
$SU(2) \times U(1)_{Y} \ra U(1)_{Q}$.

To quantize around the classical vacuum, write $\phi  = v +
\phi '$, where $\phi '$ are quantum fields with zero vacuum
expectation value. To display
the physical particle content it is useful to rewrite the four \her\
components of $\phi '$ in terms of a new set of variables using  the
Kibble transformation~\cite{Kibble:1967sv}: \beq \phi  = \frac{1}{\sqrt{2}}
e^{i \sum \xi^i L^i} \left( \begin{array}{c} 0 \\ \nu + H \end{array}
\right). \label{eqch29} \eeq $H$ is a \her\ field which will turn out to be
the physical Higgs
scalar. If we
had been dealing with a spontaneously broken global symmetry the three \her\
fields $\xi^i$ would be the massless pseudoscalar Nambu-Goldstone bosons~\cite{Nambu:1960xd,Nambu:1961tp,Goldstone:1961eq,Goldstone:1962es}
that are necessarily associated with broken symmetry generators.  However, in
a            gauge theory they disappear from the physical spectrum. To see
this      it is useful to go to the unitary gauge
\beq
\phi  \ra \phi ' = e^{-i \sum \xi^i L^i} \phi  =
\frac{1}{\sqrt{2}} \left( \begin{array}{c} 0 \\ \nu + H \end{array}
\right), \label{eqch30} \eeq
                           in which the Goldstone bosons
disappear.                          In this gauge, the scalar
covariant kinetic energy term takes the simple form
\begin{eqnarray} (D_\mu \phi )^{\dag} D^\mu \phi  & = & \frac{1}{2} (0\;
\nu) \left[ \frac{g}{2} \tau^i W^i_\mu + \frac{g'}{2} B_\mu \right]^2
\left( \begin{array}{c} 0 \\ \nu \end{array} \right) 
+ H \text{ terms} \nonumber \\
& \ra & M^2_W W^{+\mu} W^-_\mu +
\frac{M_Z^2}{2} Z^\mu Z_\mu + H \text{ terms}, \label{eqch31}
\end{eqnarray}
where the kinetic energy and gauge interaction terms of the physical $H$
particle have been omitted. Thus,
 spontaneous symmetry breaking generates mass terms for the
$W$ and $Z$ gauge bosons
\begin{eqnarray} W^{\pm}&=&\frac{1}{\sqrt{2}} (W^1 \mp i W^2)\nonumber\\
Z &=& -\sin \theta_W B + \cos \theta_W W^3.\label{eqch33} \end{eqnarray}
The photon field
\beq A = \cos \theta_W B + \sin \theta_W W^3 \label{eqch34}
\eeq
remains massless.  The masses are
\beq M_W = \frac{g \nu}{2} \label{eqch34b} \eeq
and
\beq M_Z = \sqrt{g^2 + g^{\prime 2} } \frac{\nu}{2} = \frac{M_W}{\cos
\theta_W}, \label{eqch35} \eeq
where the weak angle is defined by 
\beq \tan\theta_W\equiv \frac{g'}{g} \ \Rightarrow\ \sin^2 \theta_W= 1-\frac{M_W^2}{M_Z^2}.\label{eqch35a} \eeq
One can think of the generation of masses as due to the fact that the
$W$ and $Z$ interact constantly with the condensate of scalar fields and
therefore acquire masses, in analogy with a photon propagating through a
plasma. The Goldstone boson has
disappeared from the theory  but has reemerged as the longitudinal degree
of freedom of a massive vector particle.

It will be seen below that  $G_F/\sqrt{2} \sim g^2/8 M^2_W$, where $G_F
= 1.16637(5) \times \p{-5}$ GeV$^{-2}$ is the Fermi constant determined by the
muon lifetime. The weak scale $\nu$ is therefore 
\beq \nu = 2M_W/g \simeq
(\sqrt{2} G_F)^{-1/2} \simeq 246  \text{ GeV}. \label{3-45a} \eeq
Similarly, $g = e/\sin \theta_W$, where $e$ is the electric charge of the
positron. Hence, to lowest order 
\beq
M_W = M_Z \cos \theta_W \sim \frac{(\pi \alpha/\sqrt{2} G_F)^{1/2}}{
\sin \theta_W}, \label{wztree} \eeq
where $\alpha \sim 1/137.036$ is the fine structure constant. Using $\sin^2
\theta_W \sim 0.23$ from neutral current scattering, one expects $M_W \sim
78$ GeV, and $M_Z \sim 89$ GeV. (These predictions are increased by
$\sim (2-3)$ GeV by loop corrections.)  
          The $W$ and $Z$ were discovered at CERN by the UA1~\cite{Arnison:1985jk}
           and UA2~\cite{Ansari:1987vg}
groups in 1983.  Subsequent measurements of their masses and other
properties have been in excellent agreement with the standard model
expectations (including the higher-order corrections)~\cite{Amsler:2008zz}.
The current values are
\beq M_W = 80.398 \pm 0.025  \text{ GeV}, \qquad  M_Z = 91.1876\pm 0.0021 \text{ GeV}.
\eeql{eq7:ssb17}

\section{The Higgs and Yukawa Interactions}\label{higgssection}
\begin{figure}[htbp]
\begin{center}
\includefigure{0.55}{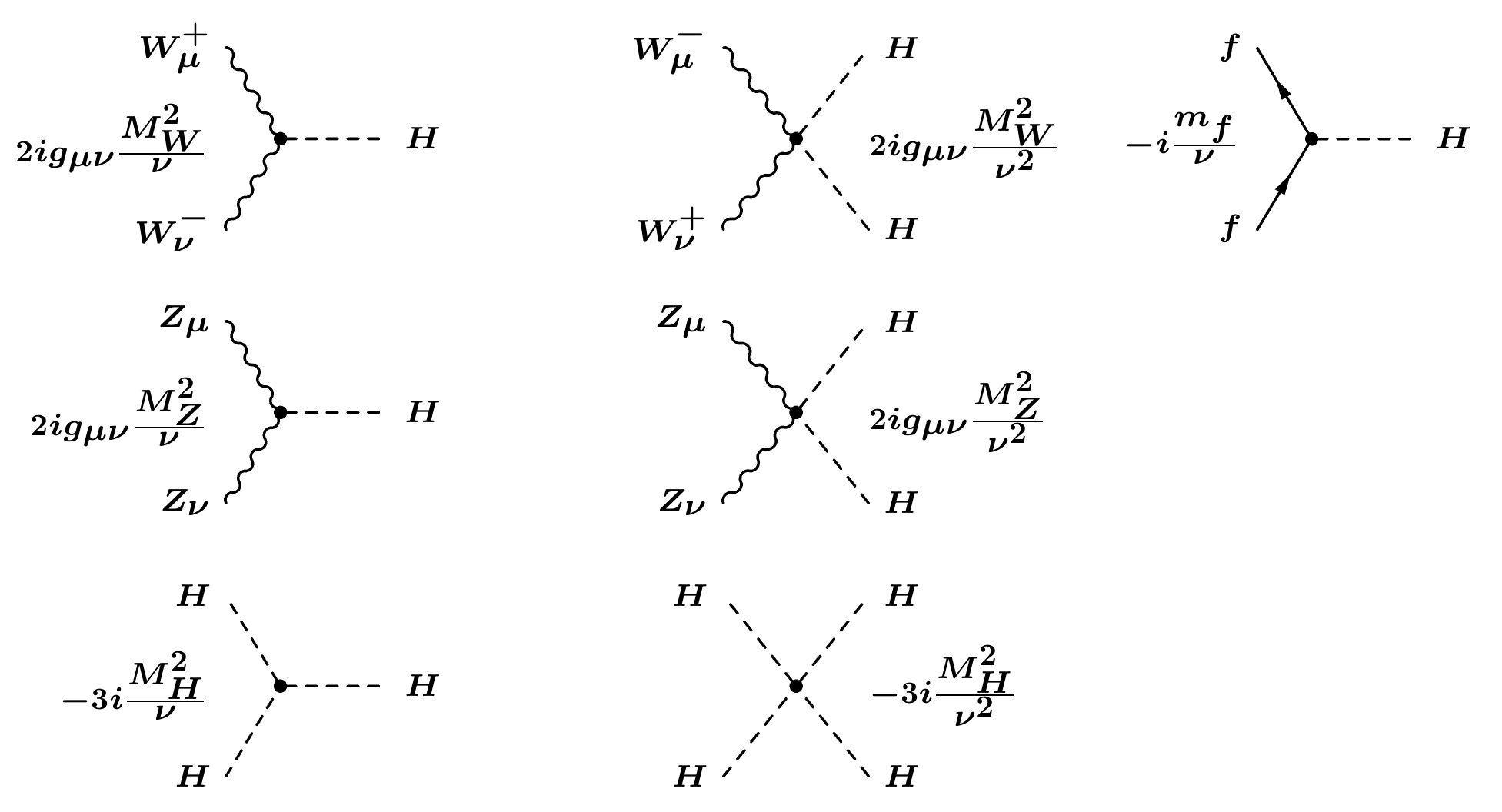}
\end{center}
\caption{Higgs interaction vertices in the standard model.}
\label{higgsints}
\end{figure}

The full Higgs part of \lag\ is
\beq
\begin{split}
\lag_\vp &=(D^\mu \vp)^{\dag} D_\mu \vp - V(\vp) \\
&= M_W^2 W^{\mu+}W_\mu^-
\left( 1+\frac{H}{\nu}\right)^2 
+ \oh M_Z^2 Z^\mu Z_\mu \left( 1+\frac{H}{\nu}\right)^2\\
&+ \oh \left(\partial_\mu H \right)^2 - V(\vp).
\end{split}\eeql{eq7:ssb18}
The second line
includes the $W$ and $Z$ mass terms and also the $ZZH^2$, $W^+W^-H^2$
and the induced $ZZH$ and $ W^+W^-H$ 
interactions, as shown in Table \ref{ch7_smrules} and Figure \ref{higgsints}.
The last line includes the canonical Higgs kinetic energy term and the potential.
\begin{table}[htdp]
\tbl{Feynman rules for the gauge and Higgs interactions after SSB,
taking combinatoric factors into account.
The momenta and quantum numbers flow into the vertex. 
 Note the dependence on $M/\nu$ or $M^2/\nu$.}
{\begin{tabular}{llll} \toprule
$ W^+_{\mu}W_\nu^- H$: & $\oh  i g_{\mu\nu} g^2\nu=2 i g_{\mu\nu}\frac{M_W^2}{\nu}$ \
&$ W^+_{\mu}W_\nu^- H^2$: & $\oh  i g_{\mu\nu} g^2=2 i g_{\mu\nu}\frac{M_W^2}{\nu^2}$\vspace*{3pt}\\
$  Z_{\mu}Z_\nu H$: & $ \frac{ i g_{\mu\nu} g^2\nu}{2\cos^2\theta_W}=2 i g_{\mu\nu}\frac{M_Z^2}{\nu}$ \
&$ Z_{\mu}Z_\nu H^2$: & $\frac{ i g_{\mu\nu} g^2}{2\cos^2\theta_W}=2 i g_{\mu\nu}\frac{M_Z^2}{\nu^2}$\vspace*{3pt}\\
$H^3$: & $-6i \lambda\nu=-3i\frac{M_H^2}{\nu}$ \
&$H^4$: & $-6i \lambda=-3i\frac{M_H^2}{\nu^2}$ \vspace*{3pt}\\
$H\bar ff$: & $-ih_f=-i\frac{m_f}{\nu} $ & 
 &   \vspace*{3pt}\\
 \multicolumn{3}{l}{$W^+_\mu (p) \gamma_\nu(q)W^-_\sigma(r) \qquad ie\,  {\cal C}_{\mu\nu\sigma}(p,q,r)$} & \vspace*{3pt} \\
\multicolumn{3}{l}{$W^+_\mu (p) Z_\nu(q)W^-_\sigma(r) \qquad  i\frac{e}{\tan \theta_W}  {\cal C}_{\mu\nu\sigma}(p,q,r)$} & \vspace*{3pt} \\
\multicolumn{3}{l}{$W^+_\mu W^+_\nu W^-_\sigma W^-_\rho \qquad i\frac{e^2}{\sin^2 \theta_W}  {\cal Q}_{\mu\nu\rho\sigma} $} &   \vspace*{3pt} \\
\multicolumn{3}{l}{$W^+_\mu Z_\nu \gamma_ \sigma W^-_ \rho \qquad- i\frac{e^2}{\tan \theta_W}  {\cal Q}_{\mu\rho\nu\sigma}$} &   \vspace*{3pt} \\
\multicolumn{3}{l}{$W^+_\mu Z_\nu Z_ \sigma W^-_ \rho \qquad - i\frac{e^2}{\tan^2 \theta_W}  {\cal Q}_{\mu\rho\nu\sigma}$}&   \vspace*{3pt} \\
\multicolumn{3}{l}{$W^+_\mu \gamma_\nu \gamma_ \sigma W^-_ \rho \qquad- ie^2 {\cal Q}_{\mu\rho\nu\sigma}$}&   \vspace*{3pt} \\
\multicolumn{4}{l}{${\cal C}_{\mu\nu\sigma}(p,q,r)\equiv  g_{\mu \nu} ( q - p )_{\sigma} 
 +g_{\mu \sigma} (p - r)_{\nu} + g_{\nu \sigma} (r - q)_{\mu}$}   \vspace*{3pt} \\
\multicolumn{4}{l}{${\cal Q}_{\mu\nu\rho\sigma} \equiv 2 g_{\mu\nu}g_{\rho\sigma}- g_{\mu \rho}g_{\nu\sigma}
- g_{\mu \sigma}g_{\nu \rho} $}   \vspace*{3pt} \\ \botrule
\end{tabular}}
\label{ch7_smrules}
\end{table}

After  symmetry breaking   the Higgs potential in unitary gauge becomes
\beq V(\phi ) = - \frac{\mu^4}{4\lambda} - \mu^2 H^2 + \lambda \nu
H^3 + \frac{\lambda}{4} H^4.\label{eqch37} \eeq
The first term in the Higgs potential $V$ is a constant, $\vev{V(\nu)}=-\mu^4/4\lambda$.
It reflects the fact that $V$ was defined so that $V(0)=0$, and therefore $V<0$ at the minimum.
Such a constant term is irrelevant to physics in the absence of gravity,  but will
be seen in Section \ref{gravity} to be one of the most serious problems of the
SM when gravity is incorporated because it acts like a cosmological constant
much larger (and of opposite sign) than is allowed by observations.
The third and fourth terms in $V$  represent the induced cubic and
quartic interactions of the Higgs scalar, shown in  Table \ref{ch7_smrules} and Figure \ref{higgsints}.

The second term in $V$ represents a (tree-level) mass 
\beq M_H = \sqrt{-2\mu^2} = \sqrt{2 \lambda} \nu , \label{eqch38} \eeq
for the Higgs boson.
The weak scale is given in (\ref{3-45a}), but the quartic Higgs coupling
$\lambda$ is unknown, so $M_H$ is not predicted. A priori, $\lambda$
could be anywhere
in the range $ 0 \leq \lambda < \infty $. 
 There is an experimental lower limit $M_H \gtrsim
114.4$~GeV at 95\% cl from LEP~\cite{Barate:2003sz}.  Otherwise, the decay $Z \ra Z^* H$ would have
been observed. 
 
  There are also plausible theoretical
limits. If $\lambda > \mathcal{O} (1)$
the theory becomes strongly coupled
$(M_H > \mathcal{O} (1$~TeV)).  There is not really anything wrong with strong
coupling a priori.  However, there are fairly convincing triviality
limits, which basically say that the running
quartic coupling would become infinite within the domain of validity
of the theory if $\lambda$ and therefore $M_H$ is too large.  If one
requires the theory to make sense to infinite energy, one runs
into problems\footnote{This is true for a  pure $\lambda H^4$ theory.
The presence of other interactions may eliminate the problems for
small $\lambda$.}
 for any $\lambda$.
However, one only needs for  the theory to hold up to the next
mass scale $\Lambda$, at which point the standard model breaks down. In that
case~\cite{Cabibbo:1979ay,Gunion:425736,Hambye:1997ax}, 
\beq
M_H < \left\{ \begin{array}{l} \mathcal{O} (180) \;GeV, \; \Lambda \sim M_P\\
                               O (700) \;GeV, \; \Lambda \sim 2M_H.
\end{array} \right. \label{eqch39} \eeq
The more stringent limit of $O(180)$~GeV obtains for $\Lambda$ of order of
the Planck scale $M_P = G_N^{-1/2} \sim 10^{19}$~GeV.  If one makes
the less restrictive assumption that the scale $\Lambda$ of new
physics can be small, one obtains a weaker limit.  Nevertheless, for
the concept of an elementary Higgs field to make sense one should
require that the theory be valid up to something of order of $2M_{H}$,
which implies that $M_H < O$(700)~GeV.  
These estimates rely on perturbation theory, which breaks down for large $\lambda$. However, they
can be justified by nonperturbative lattice calculations~\cite{Hasenfratz:1987eh,Kuti:1987nr,Luscher:1988uq}, which suggest an absolute upper limit
of $650-700$ GeV. There are also comparable upper bounds
from the validity of unitarity at the tree level~\cite{Lee:1977eg}, and {\em lower} limits from vacuum stability~\cite{Cabibbo:1979ay,Altarelli:1994rb,Casas:1996aq,Isidori:2001bm}.
The latter again depends on the scale $\Lambda$, and requires $M_H \gtrsim$ 130 GeV
for $\Lambda=M_P$ (lowered to $\sim 115$ GeV if one allows a sufficiently long-lived metastable vacuum~\cite{Casas:1996aq,Isidori:2001bm}), with a weaker constraint for lower $\Lambda$.

The Yukawa interaction  in the unitary gauge  becomes
\begin{eqnarray} -\lag_{  Yuk} &\ra& \sum^F_{m,n = 1} \bar{u}^0_{mL}
\Gamma^u_{mn}\left( \frac{\nu + H}{\sqrt{2}} \right) u^0_{mR} + (d,
e, \nu) \text{ terms}\; + \;\; h.c. \nonumber \\
&=& \bar{u}^0_L \left( M^u + h^u H \right) u^0_R + (d, e, \nu) \text{ terms}\; + h.c., \label{eqch43} \end{eqnarray}
where in the second form
$u^0_L = \left( u^0_{1L} u^0_{2L} \cdots u^0_{FL} \right)^T$
is an $F$-component column vector, with a similar definition for
$u_R^0$.   $M^u$ is an $F\times F$
fermion mass matrix
$M^u_{mn} = \Gamma^u_{mn} {\nu}/{\sqrt{2}}$
induced by spontaneous symmetry
breaking, and  $ h^u = M^u/\nu = {g M^u}/{2 M_W}$ is the Yukawa
coupling matrix.
 
In general $M$ is not diagonal, \her, or symmetric. To
identify the physical  particle content it is necessary to
diagonalize $M$ by separate unitary transformations $A_L$ and $A_R$ on the
left- and right-handed fermion fields.  (In the special case that $M^u$ is
\her\ one can take $A_L = A_R$).  Then, \beq A_L^{u \dag} M^u A^u_R = M^u_D
= \left( \begin{array}{ccc} m_u & 0 & 0 \\ 0 & m_c & 0 \\ 0 & 0 & m_t
\end{array} \right) \label{eqch44} \eeq
is a diagonal matrix with eigenvalues equal to the physical masses of
the charge $\frac{2}{3}$ quarks\footnote{
 From \refl{eqch44} and its conjugate one has
$\hat A^{u\dagger}_LM^uM^{u\dagger} \hat A^u_L=\hat A^{u\dagger}_RM^{u\dagger}M^u \hat A^u_R
=M^{u2}_D$.  But $MM^\dagger$ and $M^\dagger M$ are Hermitian, so $A_{L,R}$ can then be constructed by elementary techniques,
up to overall phases that can be chosen to make the mass eigenvalues real and positive, and to
remove unobservable phases from the weak charged current.}. Similarly, one
diagonalizes the down quark, charged lepton, and neutrino mass matrices by
\beq\begin{split}
A^{d\dag}_L M^d A^d_R &= M^d_D \\
A^{e\dag}_L M^e A^e_R &= M^e_D \\
A^{\nu\dag}_L M^\nu A^\nu_R &= M^\nu_D.
\end{split} \label{eqch45} \eeq
In terms of these unitary matrices we can define mass eigenstate
fields
$u_L = A_L^{u\dag} u_L^0 = (u_L\ c_L\ t_L)^T$,  with analogous definitions
for $u_R = A_R^{u\dag} u_R^0$,
$d_{L,R} = A_{L,R}^{d\dag} d_{L,R}^0$, $
e_{L,R} = A_{L,R}^{e\dag} e_{L,R}^0$, and $\nu_{L,R} = A_{L,R}^{\nu\dag} \nu_{L,R}^0$.
Typical estimates of the  quark masses
are~\cite{Gasser:1982ap,Amsler:2008zz} 
$m_u\sim 1.5-3$ MeV, $m_d\sim 3-7$  MeV, $m_s\sim 70-120$ MeV, $m_c\sim
1.5-1.8$ GeV, $m_b \sim 4.7-5.0$ GeV, and
$ m_t =170.9\pm 1.8$ GeV. These are
the current masses: for QCD their effects are identical to bare masses in
the QCD \lagl.  They should not be confused with the constituent masses of
order 300~MeV generated by the spontaneous breaking of chiral symmetry in
the strong interactions.  Including QCD renormalizations, the $u, \; d$, and
$s$  masses are running masses evaluated at
2 GeV$^2$, while $m_{c,b,t}$  are pole masses.
 
So far we have only allowed for ordinary Dirac mass terms
of the form $\bar \nu^0_{mL} \nu^0_{nR}$ for the neutrinos, which can be generated by the
ordinary Higgs mechanism. Another possibility are lepton number violating {Majorana masses},
which require an extended Higgs sector or higher-dimensional operators. It is not clear yet whether Nature utilizes Dirac masses, Majorana masses, or both\footnote{For reviews, see~\refcite{GonzalezGarcia:2002dz,Langacker:2005pfa,Mohapatra:2005wg,GonzalezGarcia:2007ib}.}. What is known, is that the neutrino mass eigenvalues are tiny compared to the other masses, $\lesssim {\cal O}(0.1)$ eV,
and most experiments are insensitive to them.
In describing such processes, one can ignore $\Gamma^\nu$, and the $\nu_R$ effectively decouple. Since $M^\nu\sim 0$ the three mass eigenstates are effectively degenerate with eigenvalues $0$, and 
the eigenstates are arbitrary.   That is, there is nothing to distinguish them except their weak interactions,
so we can simply define
$\nu_e,\; \nu_\mu,\; \nu_\tau$  as the weak interaction partners of
the $e, \; \mu$, and $\tau$,
which is equivalent to choosing $ A_L^{\nu}\equiv A_L^e$ so that $\nu_L = A_L^{e\dag} \nu_L^0$.
 Of course, this is not appropriate for physical processes, such as oscillation
experiments, that {\em are} sensitive to the masses or mass differences.

In terms of the mass eigenstate fermions,
\beq -\lag_{  Yuk} =  \sum_i m_i\bar{\psi}_i  \psi_i
\left(1+\frac{g}{2M_W}H\right)= \sum_i m_i\bar{\psi}_i  \psi_i
\left(1+\frac{H}{\nu}\right). \label{eqch48} \eeq   The
 coupling of the physical Higgs boson to the $i^{  th}$ fermion
is $gm_i/2M_W$, which is very small except for the top quark.  The coupling
is flavor-diagonal in the minimal model: there is just one Yukawa
matrix for each type of fermion, so the mass and Yukawa matrices are 
diagonalized by
the same transformations.
 In generalizations in
which more than one Higgs doublet couples to each type of fermion there will
in general be flavor-changing Yukawa interactions involving the
physical neutral Higgs fields~\cite{Glashow:1976nt}.  
There are stringent limits on such
couplings; for example, the $K_L - K_S$ mass difference
implies $ h/M_H < 10^{-6}\, \text{GeV}^{-1}$, where $h$ is the $\bar{d} s$
Yukawa coupling~\cite{Gaillard:1974hs,Langacker:1991uv,Nir:2007xn}.

\section{The Gauge Interactions}

The major quantitative tests of the electroweak standard model involve
the gauge interactions of fermions and the properties of the gauge
bosons.  The charged current weak interactions of the Fermi theory and
its extension to the intermediate vector boson theory\footnote{For a historical sketch, see~\refcite{Langacker:1991uv}.} are incorporated
into the standard model, as is quantum electrodynamics.  The theory
successfully predicted the existence and properties of the weak
neutral current. 
In this section I summarize the structure of the gauge interactions of fermions.

\subsection{The Charged Current}\label{wccints}

\begin{figure}[htbp]
\begin{center}
\includefigure{0.65}{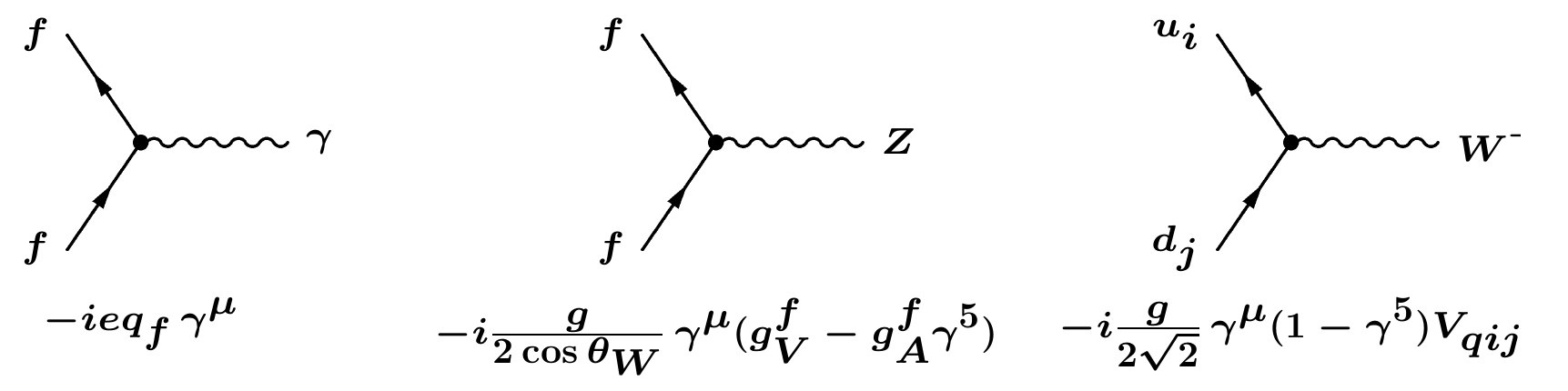}
\end{center}
\caption{The fermion gauge interaction vertices in the standard electroweak model. $g_V^f \equiv t^3_{fL} - 2 \sin^2 \theta_W q_f$ and
$g_A^f \equiv  t^3_{fL}$, where $t^3_{uL} = t^3_{\nu L}=+\oh$, while $t^3_{dL} = t^3_{eL}=-\oh$.
The $\bar d_j u_i W^-$ vertex is the same as for $\bar u_i d_j W^+$ except $V_{qij} \ra \left( V^\dag_{q} \right)_{ji}=V^\ast_{qij}$.
The lepton-$W^\pm$ vertices are obtained from the quark ones by $u_i\ra \nu_i$, $d_j \ra e^-_j$,
and $V_q\ra V_\ell$.}
\label{fermiongauge}
\end{figure}

The interaction of the $W$ bosons to fermions is given by
\beq \lag = - \frac{g}{2\sqrt{2}} \left( J^\mu_W W^-_\mu + J^{\mu\dag
}_W W^+_\mu \right), \label{eqch49} \eeq
where the weak charge-raising current is
\beq
\begin{split}
J_W^{\mu \dag} &= \sum^F_{m=1} \left[ \bar{\nu}_m^0 \gamma^\mu (1 -
\gamma^5) e^0_m + \bar{u}_m^0 \gamma^\mu (1-\gamma^5) d^0_m \right]
      \\
&= (\bar{\nu}_e \bar{\nu}_\mu \bar{\nu}_\tau ) \gamma^\mu (1 -
\gamma^5)V_\ell \left( \begin{array}{c} e^- \\ \mu^- \\ \tau^- \end{array}
\right) + (\bar{u}\; \bar{c}\; \bar{t}) \gamma^\mu (1 - \gamma^5) V_q
\left( \begin{array}{c} d \\ s \\ b \end{array} \right).
\end{split} \label{eqch50}   \eeq
$J^{\mu \dag}_W$ has a $V-A$ form, i.e., it violates parity and charge
conjugation maximally.  
The fermion gauge vertices are shown in Figure \ref{fermiongauge}.

The mismatch between the unitary
transformations relating the weak and mass eigenstates for the up and
down-type quarks leads to the presence of the $F \times F$ unitary matrix $V_q \equiv
A^{u \dag}_L A^d_L $  in the current.  This is the
Cabibbo-Kobayashi-Maskawa (CKM) matrix~\cite{Cabibbo:1963yz,Kobayashi:1973fv}, which is ultimately due
to the mismatch between the weak and Yukawa interactions.  For $F = 2$
families $V_q$ takes the familiar form\footnote{An arbitrary $F \times F$ unitary matrix
involves $F^2$ real parameters. In this case $2F-1$ of them are unobservable relative phases in the fermion mass eigenstate fields, leaving $F(F-1)/2$ rotation angles and $(F-1)(F-2)/2$ observable $CP$-violating phases.
There are an additional $F-1$ Majorana phases in $V_\ell$ for Majorana neutrinos.}
\beq V_{Cabibbo} = \left( \begin{array}{cc} \cos \theta_c & \sin\theta_c \\ -
\sin \theta_c & \cos \theta_c \end{array} \right), \label{eqch51} \eeq
where $\sin \theta_c \simeq 0.22$ is the Cabibbo angle.  This form
gives a good zero$^{  th}$-order approximation to the weak
interactions of the $u, d,s $ and $c$ quarks; their coupling to the
third family, though non-zero, is very small.
Including these couplings, the 3-family CKM matrix is
\beq
V_{CKM} = \left( \begin{array}{ccc} V_{ud} & V_{us} & V_{ub} \\
V_{cd} & V_{cs} & V_{cb} \\ V_{td} & V_{ts} & V_{tb} \end{array}\right)
\sim \left( \begin{array}{ccc} 1 & \lambda & \lambda^3 \\
\lambda & 1 & \lambda^2 \\ \lambda^3 & \lambda^2 & 1 \end{array} \right),
 \label{eq330.10} \eeq
where the $V_{ij}$ may involve a $CP$-violating phase.
The second form, with  $\lambda=\sin \theta_c$, is an easy to remember 
approximation to the observed magnitude of each element~\cite{Wolfenstein:1983yz}, which displays a suggestive but not well understood  hierarchical structure.
These are order of magnitude only; each element may be multiplied by a phase and a coefficient of ${\cal O}(1)$.

$ V_{\ell}\equiv A^{\nu\dagger}_L A^e_L$ in \refl{eqch50} is the analogous leptonic mixing matrix. It is critical for describing
neutrino oscillations and other processes sensitive to neutrino masses.
However, for processes for which the neutrino masses are negligible we can
effectively set  $ V_{\ell}=I$ (more precisely, $ V_{\ell}$ will only enter such processes in the
combination $ V_{\ell}^\dagger V_{\ell}=I$, so it can be ignored).

\begin{figure}
\centering
\includegraphics*[scale=0.6]{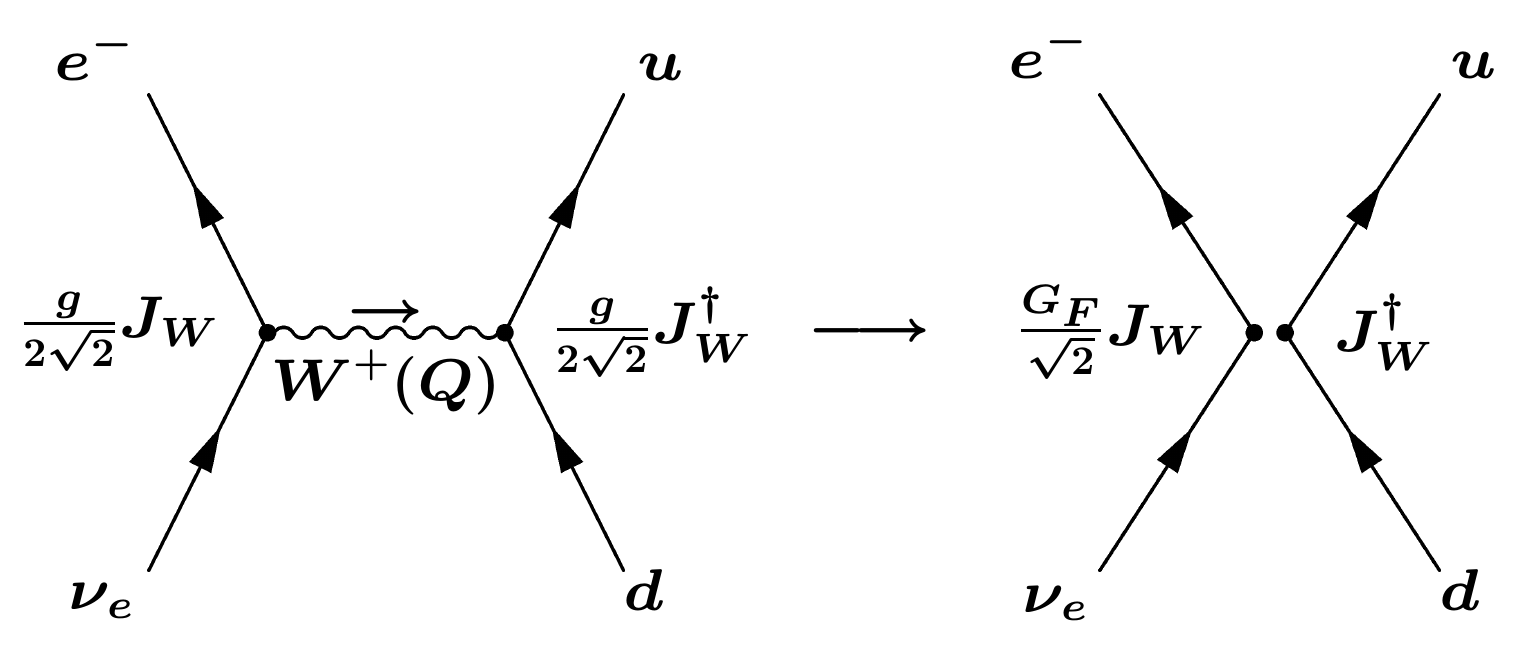}
\caption{A weak interaction mediated by the exchange of a $W$ and the
effective four-fermi interaction that it generates if the
four-momentum transfer $Q$ is sufficiently small.}
\label{figure12b}
\end{figure}
 
The interaction between fermions mediated by the exchange of a $W$ is
illustrated in Figure~\ref{figure12b}.  In the limit $|Q^2| \ll M_W^2$
the momentum term in the $W$ propagator can be neglected, leading to
an effective zero-range (four-fermi) interaction
\beq -\lag^{cc}_{  eff} = \frac{G_F}{\sqrt{2}} J^\mu_W J^{\dag}_{W\mu},
\label{eqch52} \eeq
where the Fermi constant is identified as
\beq \frac{G_F}{\sqrt{2}} \simeq \frac{g^2}{8M_W^2} =
\frac{1}{2\nu^2}.
\label{eqch53} \eeq
Thus, the Fermi theory is an approximation to the standard model valid
in the limit of small momentum transfer.
From the muon lifetime,  $G_F
= 1.16637(5) \times 10^{-5}$~GeV$^{-2}$, which implies that the weak
interaction scale defined by the VEV of the Higgs field is $\nu =
\sqrt{2} \langle 0| \phi ^0 |0 \rangle \simeq 246$~GeV.
 
The charged current weak interaction as described by (\ref{eqch52})
has been
successfully tested in a large variety of weak decays~\cite{Commins:100123,Renton:206324,Langacker:268088,Amsler:2008zz},
including $\beta$, $K$, hyperon, heavy quark, $\mu$, and $\tau$
decays.  In particular, high precision measurements of $\beta$, $\mu$, and
$\tau$ decays are a sensitive probe of extended gauge groups involving
right-handed currents and other types of new physics, as is described in
the chapters by Deutsch and Quin; Fetscher and
Gerber; and Herczeg in~\refcite{Langacker:268088}. Tests of the unitarity
of the CKM matrix are important in searching for the presence of fourth
family or exotic fermions and for new interactions~\cite{Czarnecki:2004cw}. The standard theory has also
been successfully probed in neutrino scattering processes such as $ \nu_\mu
e \ra \mu^- \nu_e, \nu_\mu n \ra \mu^- p, \nu_\mu N \ra \mu^- X $. It
works so well that the charged current neutrino-hadron interactions are used more as a probe
of the structure of the hadrons and QCD than as a test of the weak
interactions.
 
\begin{figure}[htbp]
\begin{center}
\includefigure{0.55}{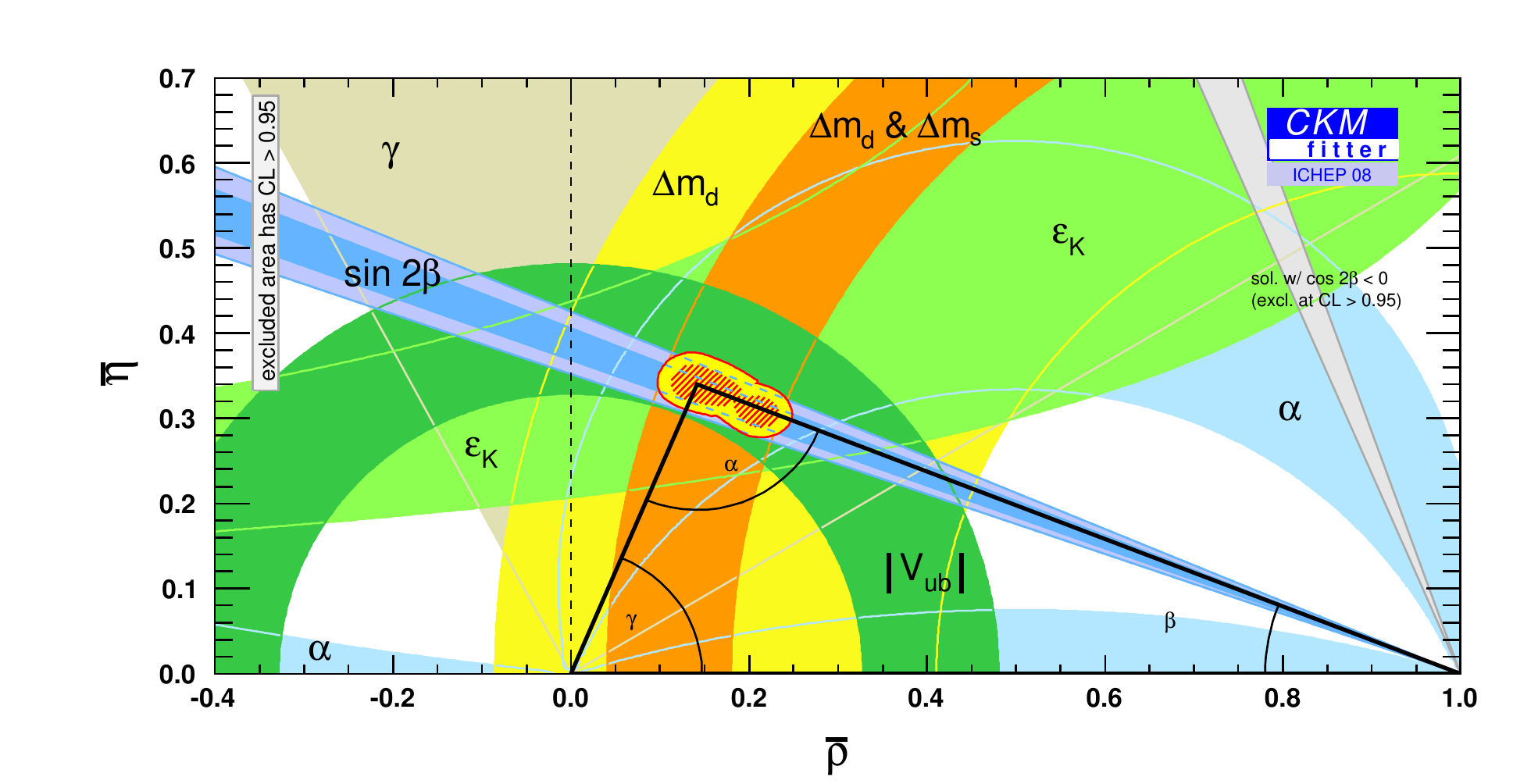}
\end{center}
\caption{The unitarity triangle, showing the consistency of various \CP-conserving and \CP-violating
observables from the $K$ and $B$ systems. $\bar \rho$ and $\bar \eta$
are the same as $\rho$ and $\eta$ up to higher order corrections,
where $\rho-i\eta=V_{ub}/(V_{cb}V_{us})$. Plot courtesy of the CKMfitter group~\cite{Charles:2004jd}, {\tt http://ckmfitter.in2p3.fr}.}
\label{unitaritytri}
\end{figure}

Weak charged current effects have also been observed in higher orders,
such as in $K^0-\bar K^0$, $D^0-\bar D^0$, and $B^0-\bar B^0$ 
mixing, and in \CP\ violation in $K$ and $B$ decays~\cite{Amsler:2008zz}.
For these higher order processes the full theory must
be used because large momenta occur within the loop integrals.
An example of the consistency between theory and experiment is shown in Figure \ref{unitaritytri}.

\subsection{QED}

The standard model incorporates all of the (spectacular) successes of quantum
electrodynamics (QED), which is based on the $U(1)_{Q}$ subgroup that
remains unbroken after spontaneous symmetry breaking.  The relevant
part of the \lagl\ density is
\beq \lag = - \frac{gg'}{\sqrt{g^2 + g^{\prime 2}}} J^\mu_{Q} ( \cos
\theta_W B_\mu + \sin\theta_W W^3_\mu), \label{eqch54} \eeq
where the linear combination of neutral gauge fields is just the
photon field $A_\mu$.  This reproduces the QED interaction provided
one identifies the combination of couplings
\beq e = g \sin \theta_W \label{eqch55} \eeq
as the electric charge of the positron, where $\tan  \theta_W \equiv
g'/g$. 
The electromagnetic current is given by
\beq \begin{split}
J^\mu_{Q} &=  \sum^F_{m=1} \left[ \frac{2}{3} \bar{u}^0_m
\gamma^\mu u^0_m - \frac{1}{3} \bar{d}^0_m \gamma^\mu d_m^0 -
\bar{e}^0_m \gamma^\mu e^0_m \right] \\
           &= \sum^F_{m=1} \left[ \frac{2}{3} \bar{u}_m
\gamma^\mu u_m - \frac{1}{3} \bar{d}_m \gamma^\mu d_m -
\bar{e}_m \gamma^\mu e_m \right].
\end{split}
\label{eqch57}\eeq
It takes the same form when written in terms of either weak
or mass eigenstates  because all fermions which mix with each
other have the same electric charge.
 Thus, the electromagnetic current is
automatically flavor-diagonal.
 
Quantum electrodynamics is the most successful theory in physics when judged in terms of the theoretical and
experimental precision of its tests. A detailed review is given in~\refcite{Kinoshita:213617}. The classical
atomic tests of QED, such as the Lamb shift, atomic hyperfine splittings, muonium ($\mu^+e^-$ bound states), and
positronium ($e^+e^-$ bound states) are reviewed in~\refcite{Karshenboim:2005iy}. The 
most precise determinations of $\alpha$ and the other physical constants are surveyed in~\refcite{Mohr:2008fa}.
High energy tests are described in~\refcite{Wu:1984ik,Kiesling:113108}.
The currently most precise measurements of  $\alpha$ are compared in Table \ref{ch2_alpha}.
The approximate agreement of these determinations, which involves the calculation of  the electron anomalous magnetic moment $a_e=(g_e-2)/2$ to high order, validates not only
QED but the entire formalism of gauge invariance and renormalization theory.
Other basic predictions of gauge invariance (assuming it is not spontaneously broken, which would lead
to electric charge nonconservation), are that the photon mass $m_\gamma$
and its charge $q_\gamma$ (in units of $e$) should vanish. The current upper bounds
are extremely impressive~\cite{Amsler:2008zz}
\beq m_\gamma < 1\times 10^{-18} \text{ eV }, \qquad q_\gamma < 5\times 10^{-30},
\eeql{ch2:qedtest1}
based on astrophysical effects (the survival of the Solar magnetic field and limits on the dispersion of
light from pulsars). 

There is a possibly significant discrepancy between the high precision measurement of the 
anomalous magnetic moment of the muon $a_\mu^{exp}=11~659~208.0(5.4)(3.3) \times 10^{-10}$ by 
the Brookhaven 821 experiment~\cite{Bennett:2006fi}, and the theoretical expectation,
for which the purely QED part has been calculated to 4 loops and the leading 5 loop contributions
estimated (see the review by H\"ocker and Marciano in \refcite{Amsler:2008zz}).
In addition to the QED part, there are weak interaction corrections (2 loop) and hadronic vacuum polarization
and hadronic light by light scattering corrections. There is some theoretical uncertainty in the hadronic
corrections.  Using estimates of the hadronic vacuum polarization using the  measured cross section for $e^+e^-\ra$ hadrons
in a dispersion relation, one finds
\beq 
a_\mu^{SM}=  116~591~788 (58)\times 10^{-11} \ \Rightarrow \  \Delta a_\mu=a_\mu^{exp}-a_\mu^{SM}= 292(86)\times 10^{-11},
\eeql{ch2:qedtest4}
a 3.4$\sigma$ discrepancy. However, using hadronic $\tau$ decay instead, the discrepancy is reduced  to only $0.9\sigma$.
If real, the discrepancy could single the effects of new physics, such as the contributions of relatively light supersymmetric
particles. For example, the central value of the discrepancy in  \refl{ch2:qedtest4}  would be accounted for~\cite{Czarnecki:2001pv} if
\beq m_{SUSY} \sim 67 \sqrt{\tan \beta} \text{ GeV},
\eeql{ch2:qedtest5}
where $m_{SUSY}$ is the typical mass of the relevant sleptons, neutralinos, and charginos,
and $\tan \beta$ is the ratio of the expectation values of the two Higgs doublets in the theory.

\begin{table}[htdp]
\tbl{Most precise determinations of the fine structure constant $\alpha=e^2/4\pi$. 
$\Delta_e$ is defined as  $\left[ \alpha^{-1}-\alpha^{-1}(a_e) \right]\times 10^6$. Detailed descriptions and
references are given in~\refcite{Mohr:2008fa}.}
{\begin{tabular}{llll} 
Experiment            & Value of $\alpha^{-1}$  & Precision                                  & $\Delta_e$\\
\hline
$a_e=(g_e-2)/2$ & $137.035~999~683~(94)$  & $[6.9 \times 10^{-10}]$ & ~~~~~~-- \vspace*{5pt}\\ 
$h/m$ (Rb, Cs)  & $137.035~999~35~(69)$  & $[5.0 \times 10^{-9}]$  &$0.33\pm 0.69$ \vspace*{5pt}\\
Quantum Hall  & $137.036~003~0~(25)$  & $[1.8 \times 10^{-8}]$  &$-3.3\pm 2.5$ \vspace*{5pt}\\
$h/m$ (neutron) &  $137.036~007~7~(28)$  & $[2.1 \times 10^{-8}]$  &$-8.0\pm 2.8$ \vspace*{5pt}\\
$\gamma_{p,^3He}$ (J. J.) &  $137.035~987~5~(43)$  & $[3.1 \times 10^{-8}]$  &$12.2\pm 4.3$ \vspace*{5pt}\\
$\mu^+e^-$ hyperfine &  $137.036~001~7~(80)$  & $[5.8 \times 10^{-8}]$  &$-2.0\pm 8.0$ \vspace*{3pt}\\
\end{tabular}}
\label{ch2_alpha}
\end{table}

\subsection{The Neutral Current}

The third class of gauge interactions is the weak neutral
current,
 which was predicted by the $SU(2) \times U(1)$ model.
The relevant interaction is
\beq \lag = - \frac{\sqrt{g^2 + g^{\prime 2}}}{2} J^\mu_Z \left( - \sin
\theta_W B_\mu + \cos \theta_W W^3_\mu \right)=-\frac{g}{2\cos\theta_W} J^\mu_Z Z_\mu, \label{eqch59} \eeq
where the combination of neutral fields is the massive $Z$ boson field.  The
strength is conveniently rewritten as $ g/(2 \cos \theta_W)$, which
follows from $\cos \theta_W= g/\sqrt{g^2 + g^{\prime 2}}$.
 
The weak neutral current is given by
\beq \begin{split}
J^\mu_Z &= \sum_m \left[\bar{u}^0_{mL} \gamma^\mu
u^0_{mL} - \bar{d}^0_{mL} \gamma^\mu d^0_{mL} + \bar{\nu}^0_{mL}
\gamma^\mu \nu^0_{mL} - \bar{e}^0_{mL} \gamma^\mu e^0_{mL} \right] \\ & \qquad \qquad
 -2 \sin^2 \theta_W J^\mu_{Q} \\
&=\sum_m \left[ \bar{u}_{mL} \gamma^\mu u_{mL} - \bar{d}_{mL}
\gamma^\mu d_{mL} + \bar{\nu}_{mL} \gamma^\mu \nu_{mL} - \bar{e}_{mL}
\gamma^\mu e_{mL} \right] \\ & \qquad \qquad
 -2 \sin^2 \theta_W J^\mu_{Q}.
\end{split} \label{eqch60}
\eeq
Like the
electromagnetic current $J^\mu_Z$ is flavor-diagonal in the standard
model; all fermions which have the same electric charge and chirality
and therefore can mix with each other have the same $SU(2) \times U(1)$
assignments, so the form is not affected by the unitary
transformations that relate the mass and weak bases.  It was for this
reason that the GIM mechanism~\cite{Glashow:1970gm} was introduced into the
model, along with its prediction of the charm quark.  Without it the
$d$ and $s$ quarks would not have had the same $SU(2) \times U(1)$
assignments, and flavor-changing neutral currents would have
resulted. The absence of such effects is a major restriction on many
extensions of the standard model involving exotic fermions~\cite{Langacker:1988ur}.
The neutral current has two contributions. The first only involves the
left-chiral fields and is purely $V-A$.  The second is proportional to the
electromagnetic current with coefficient $\sin^2\theta_W$ and is purely
vector.  Parity is therefore violated in the neutral current interaction,
though not maximally.
 
\begin{figure}
\centering
\includegraphics*[scale=0.6]{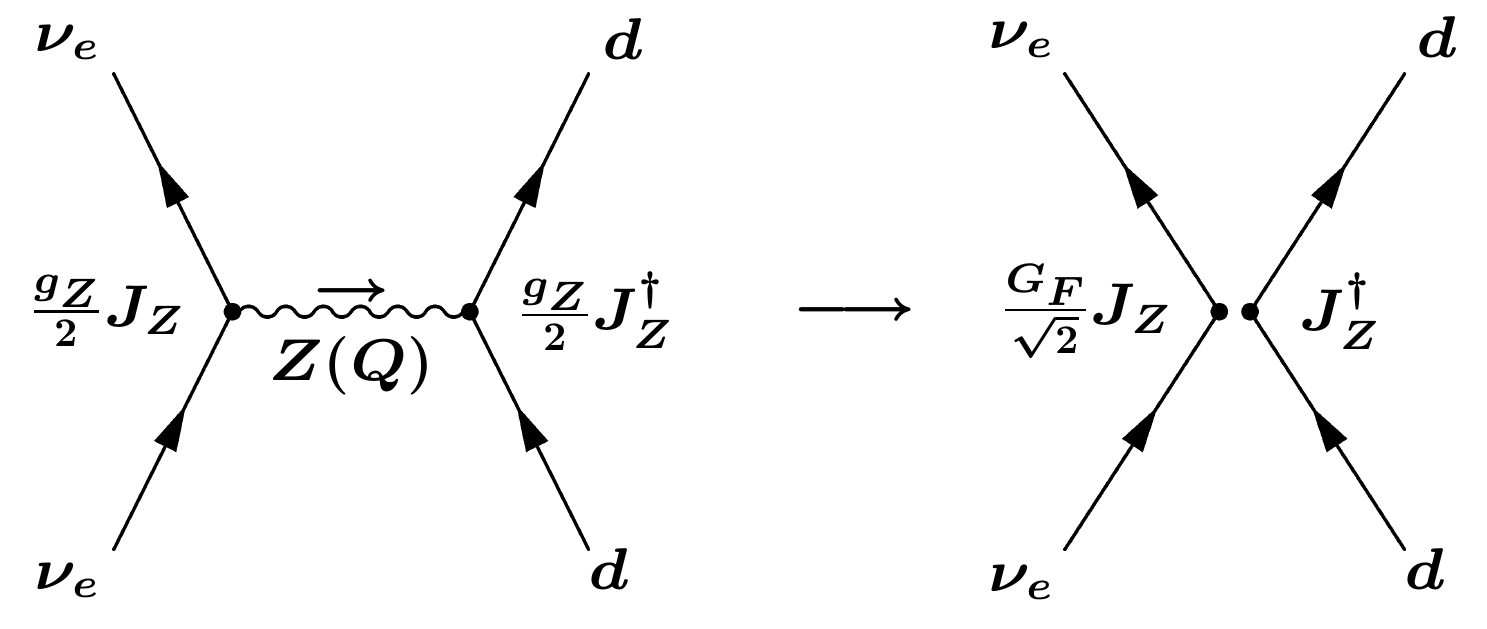}
\caption{Typical neutral current interaction mediated by the exchange
of the $Z$, which reduces to an effective four-fermi interaction in
the limit that the momentum transfer $Q$ can be neglected. $g_Z$ is defined as $\sqrt{g^2 +g^{\prime 2}}$.}
\label{figure15b}
\end{figure}
 
In an interaction
between fermions in the limit that the momentum transfer is small
compared to $M_Z$ one can neglect the $Q^2$ term in the propagator,
and the interaction reduces to an effective four-fermi interaction
\beq - \lag^{NC}_{  eff}                           =
\frac{G_F}{\sqrt{2}} J^\mu_Z J_{Z\mu}.\label{eqch61} \eeq
The coefficient is the same as in the charged case because
\beq \frac{G_F}{\sqrt{2}} = \frac{g^2}{8 M_W^2} = \frac{g^2 +
g^{\prime 2}}{8M_Z^2}. \label{eqch62} \eeq
That is, the difference in $Z$ couplings compensates the difference in
masses in the propagator.  

 The weak neutral current was discovered at
CERN in 1973 by the Gargamelle bubble chamber collaboration~\cite{Hasert:1973ff} and by HPW at
Fermilab~\cite{Benvenuti:1974xq} shortly thereafter, and since that time $Z$ exchange and $\gamma-Z$ interference  
processes have
been extensively studied in many interactions, including $ \nu e \ra
\nu e, \; \nu N \ra \nu N, \; \nu N \ra \nu X$; polarized $e^-$-hadron and $\mu$-hadron scattering; atomic parity violation; 
and in $e^+ e^-$ and $Z$-pole reactions\footnote{For reviews, see
\refcite{Kim:1980sa,Amaldi:1987fu,Costa:1987qp,Langacker:1991zr,Langacker:268088,Erler:2008ek} and the Electroweak review in \refcite{Amsler:2008zz}.
For a historical perspective, see~\refcite{Langacker:1993bu}.}. Along with the properties of the $W$ and $Z$ they
have been the primary quantitative test of the unification part of the
standard electroweak model. 

The results of these experiments have generally been in excellent agreement with the predictions of 
the SM, indicating that the basic structure is correct to first approximation and constraining the effects of possible new physics. One exception  are the recent
precise measurements of the ratios of neutral to charged current deep inelastic neutrino
scattering by the NuTeV collaboration at Fermilab~\cite{Zeller:2001hh}, with
a sign-selected beam which allowed them to minimize the effects of the $c$ threshold
in the charged current denominator. They obtained a value of $\sin^2 \theta_W=1-M_W^2/M_Z^2$ of 0.2277(16), which is 3.0$\sigma$ above the global fit value of 0.2231(3), possibly indicating new physics. However, the effect is reduced to $\sim 2\sigma$
if one incorporates the effects of the difference between the strange and 
antistrange quark momentum distributions, 
$S^- \equiv \int_0^1 {\rm d} x x [s(x) - \bar{s}(x)] = 0.00196 \pm 0.00135$,
from dimuon events, recently reported by NuTeV~\cite{Mason:2007zz}.
Other possible effects that could contribute are large isospin violation in
the nucleon sea,  next to leading
order QCD effects  and electroweak corrections, and nuclear shadowing (for a review, see~\refcite{Amsler:2008zz}).

\subsection{The $Z$-Pole and Above}
 The cross section for $e^+e^-$ annihilation is greatly enhanced near the $Z$-pole.
This allowed high statistics studies of the properties of the $Z$ at LEP (CERN) and SLC (SLAC)
in $e^- e^+ \rightarrow Z \rightarrow \ell^- \ell^+, $
  $q \bar{q},$ and $ \nu \bar{\nu}$\footnote{For reviews,
  see \refcite{LEPSLC2005ema} and the articles by D. Schaile and by A. Blondel in~\refcite{Langacker:268088}.}.
The four experiments   ALEPH, DELPHI, L3, and OPAL at LEP collected some $1.7\times 10^7$
 events at or near the $Z$-pole during the period 1989-1995. The SLD collaboration at the SLC observed some
$6 \times 10^5$ events during 1992-1998, with the lower statistics compensated by a highly polarized
$e^-$ beam with  $P_{e^-} \gtrsim 75$\%.

The basic $Z$-pole observables relevant to the precision program are:
  \begin{itemize}
   \item The  lineshape variables $M_Z,$ $ \Gamma_Z,$ and $\sigma_{peak}$.
   \item The branching ratios for $Z$ to decay into
 $e^-e^+$, $\mu^- \mu^+,$ or  $\tau^- \tau^+$; into
     $q \bar{q}$, $ c \bar{c},$ or $ b \bar{b}$; or into invisible channels such as 
   $\nu \bar{\nu}$ (allowing a determination of the number $N_\nu = 2.985 \pm 0.009$ 
       of neutrinos lighter than  $M_Z/2$).
   \item Various asymmetries, including forward-backward (FB), hadronic FB charge, polarization (LR), mixed FB-LR,
   and the polarization of produced $\tau$'s.
  \end{itemize}
 The branching ratios and FB asymmetries could be measured separately for $e$, $\mu$, and $\tau$,
 allowing tests of lepton family universality.
 
  LEP and SLC simultaneously carried out other  programs,
most notably studies and tests of QCD, and heavy quark physics.

The second phase of LEP, LEP~2, ran at CERN  from 1996-2000, with energies gradually increasing from $\sim 140$ to $\sim 209$ GeV~\cite{Alcaraz:2006mx}.
The principal electroweak results were precise measurements of the $W$ mass, as well
as its width and branching ratios;
a measurement of  $e^+ e^- \RA W^+ W^-$,  $ZZ$, and single $W$,
as a function of center of mass (CM)
energy, which tests the cancellations between diagrams that is characteristic
of a renormalizable gauge field theory, or, equivalently, probes the triple
gauge vertices;
limits on anomalous quartic gauge vertices;
measurements of various cross sections and asymmetries for
$e^+ e^- \RA f \bar{f}$ for $f=\mu^-,\tau^-,q,b$ and $c$, in reasonable
agreement with SM predictions; and
a stringent lower limit of 114.4 GeV on the Higgs mass, and even hints
of an observation at $\sim$ 116 GeV. LEP2 also 
studied heavy quark properties, tested QCD, and searched
for supersymmetric and other exotic particles.

The Tevatron $\bar p p$ collider at Fermilab has run from $\sim$1987, with a CM energy of
nearly 2 TeV. The CDF and D0 collaborations there discovered the top quark in 1995, with a mass
consistent with the predictions from the precision electroweak and $B/K$ physics observations;
have measured the $t$ mass, the  $W$ mass and decay properties, and leptonic asymmetries; carried out
Higgs searches; observed $B_s-\bar B_s$ mixing and other aspects of $B$ physics; carried out extensive QCD
tests; and searched for anomalous triple gauge couplings, heavy $W'$ and $Z'$ gauge bosons, exotic fermions, supersymmetry, and other types of new physics~\cite{Amsler:2008zz}.
The HERA $e^\pm p$ collider at DESY observed $W$ propagator and $Z$ exchange effects, searched for
leptoquark and other exotic interactions, and carried out a major program of QCD tests and structure functions studies~\cite{Wichmann:2007zf}.

The principal $Z$-pole, Tevatron, and weak neutral current experimental results are listed and compared with
the SM best fit values in Tables \ref{ch7_zpole} and \ref{ch7_nonzpole}. The $Z$-pole observations
are in excellent agreement with the SM expectations except for $A_{FB}^{0, b}$, which is the forward-backward
asymmetry in $e^-e^+ \ra b \bar b$. This could be a fluctuation or a hint of new physics (which might be expected to 
couple most strongly to the third family).
As of November, 2007, the result of the Particle Data Group~\cite{Amsler:2008zz}  global fit to all of the data was
\beq
\begin{split}
           M_H &= 77^{+28}_{-22} \mbox{ GeV}, \qquad
           m_t = 171.1  \pm 1.9  \mbox{ GeV}  \\
      \alpha_s &= 0.1217(17), \qquad
 \hat{\alpha}(M^2_Z)^{-1}  =  127.909(19), \qquad 
  \Delta \alpha_{\rm had}^{(5)} = 0.02799(14)
 \\
   \hat{s}^2_Z &= 0.23119(14), \qquad 
  \bar{s}^2_\ell = 0.23149(13), \qquad 
          s^2_W = 0.22308(30) ,
\end{split}\eeql{results}
with a good overall $\chi^2/df$ of $49.4/42$. The three values of the weak
angle $s^2$ refer to the values found using various renormalization prescriptions, viz. the
\msb, effective $Z$-lepton vertex, and on-shell
values, respectively. The latter has a larger uncertainty because of a stronger
dependence on the top mass. $ \Delta \alpha_{\rm had}^{(5)}(M_Z)$ is the hadronic contribution to the
running of the fine structure constant $ \hat{\alpha}$ in the \msb\ scheme to the $Z$-pole.

\begin{table}[htdp]
\tbl{Principal $Z$-pole observables, 
their experimental values, 
theoretical predictions using the SM parameters from the global best
fit with $M_H$ free (yielding $M_H=77^{+28}_{-22}$ GeV), pull
(difference from the prediction divided by the uncertainty), and Dev. (difference for fit
with $M_H$ fixed at 117 GeV, just above the direct search limit of 114.4 GeV), as of 11/07, from~\refcite{Amsler:2008zz}.
$\Gamma({\rm had})$, $\Gamma({\rm inv})$, and $\Gamma({\ell^+\ell^-})$ are not 
independent. }
{\begin{tabular}{lllrr} 
 Quantity & {\rm Value} & {\rm Standard Model} & {\rm Pull} & {\rm Dev.}                       \cr 
\hline \noalign{\vskip 1pt}
$M_Z$                    [GeV]   &  $91.1876 \pm 0.0021$ &  $91.1874 \pm 0.0021$  & $ 0.1$     &$ -0.1  $   \cr\noalign{\vskip 1pt}
$\Gamma_Z$               [GeV]   & $   2.4952 \pm 0.0023 $&  $   2.4968 \pm 0.0010  $&  $ -0.7     $&  $ -0.5     $ \cr\noalign{\vskip 1pt}
$\Gamma({had})$      [GeV]   & $   1.7444 \pm 0.0020 $&  $   1.7434 \pm 0.0010  $&  \hbox{---}&\hbox{---}\cr\noalign{\vskip1pt}  
$\Gamma({inv})$      [MeV]   & $ 499.0    \pm 1.5    $&  $ 501.59   \pm 0.08    $& \hbox{---}&\hbox{---}\cr\noalign{\vskip 1pt}
$\Gamma({\ell^+\ell^-})$ [MeV]   & $  83.984  \pm 0.086  $&  $  83.988  \pm 0.016   $& \hbox{---}&\hbox{---}\cr\noalign{\vskip 1pt} 
$\sigma_{had}$       [nb]    & $  41.541  \pm 0.037  $&  $  41.466  \pm 0.009   $&  $  2.0     $&  $  2.0     $ \cr\noalign{\vskip 1pt}
$R_e$                            & $  20.804  \pm 0.050  $&  $  20.758  \pm 0.011   $&  $  0.9     $&  $  1.0     $ \cr\noalign{\vskip 1pt}
$R_\mu$                          & $  20.785  \pm 0.033  $&  $  20.758  \pm 0.011   $&  $  0.8     $&  $  0.9     $ \cr\noalign{\vskip 1pt}
$R_\tau$                         & $  20.764  \pm 0.045  $&  $  20.803  \pm 0.011   $&  $ -0.9     $&  $ -0.8     $ \cr\noalign{\vskip 1pt}
$R_b$                            & $   0.21629\pm 0.00066$&  $   0.21584\pm 0.00006 $&  $  0.7     $&  $  0.7     $ \cr\noalign{\vskip 1pt}
$R_c$                            & $   0.1721 \pm 0.0030 $&  $   0.17228\pm 0.00004 $&  $ -0.1     $&  $ -0.1     $ \cr\noalign{\vskip 1pt}
$A_{FB}^{0,e}$                 & $   0.0145 \pm 0.0025 $&  $   0.01627\pm 0.00023 $&  $ -0.7     $&  $ -0.6     $ \cr\noalign{\vskip 1pt}
$A_{FB}^{0,\mu}$               & $   0.0169 \pm 0.0013 $&                        & $  0.5     $&  $  0.7     $ \cr\noalign{\vskip 1pt}
$A_{FB}^{0,\tau}$              &  $   0.0188 \pm 0.0017 $&                       & $  1.5     $&  $  1.6     $ \cr\noalign{\vskip 1pt}
$A_{FB}^{0, b}$              & $   0.0992 \pm 0.0016 $&  $   0.1033 \pm 0.0007  $&  $ -2.5     $&  $ -2.0     $ \cr\noalign{\vskip 1pt}
$A_{FB}^{0, c}$              & $   0.0707 \pm 0.0035 $&  $   0.0738 \pm 0.0006  $&  $ -0.9     $&  $ -0.7     $ \cr\noalign{\vskip 1pt}
$A_{FB}^{0, s}$              & $   0.0976 \pm 0.0114 $&  $   0.1034 \pm 0.0007  $&  $ -0.5     $&  $ -0.4     $ \cr\noalign{\vskip 1pt}
$\bar{s}_\ell^2(A_{FB}^{0,q})$ (LEP) & $   0.2324 \pm 0.0012 $&  $   0.23149\pm 0.00013 $&  $  0.8     $&  $  0.6     $ \cr\noalign{\vskip 1pt}
         {$\bar{s}_\ell^2(A_{FB}^{0,e})$} (CDF)                      & $   0.2238 \pm 0.0050 $&                        & $ -1.5     $&  $ -1.6     $ \cr\noalign{\vskip 1pt}
$A_e$   \ \  ($hadronic$)                         & $   0.15138\pm 0.00216$&  $   0.1473 \pm 0.0011  $&  $  1.9     $&  $  2.4     $ \cr\noalign{\vskip 1pt}
\phantom{$A_e$}   \ \  ($leptonic$)                                  & $   0.1544 \pm 0.0060 $&                        & $  1.2     $&  $  1.4     $ \cr\noalign{\vskip 1pt}
\phantom{$A_e$}   \ \    ($P_\tau$)                                & $   0.1498 \pm 0.0049 $&                       & $  0.5     $&  $  0.7     $ \cr\noalign{\vskip 1pt}
$A_\mu$                          & $   0.142  \pm 0.015  $&                       & $ -0.4     $&  $ -0.3     $ \cr\noalign{\vskip 1pt}
$A_\tau$   \ \     (SLD)                  & $   0.136  \pm 0.015  $&                       & $ -0.8     $&  $ -0.7     $ \cr\noalign{\vskip 1pt}
\phantom{$A_\tau$}  \ \     ($P_\tau$)                                     & $   0.1439 \pm 0.0043 $&                        & $ -0.8     $&  $ -0.5     $ \cr\noalign{\vskip 1pt}
$A_b$                            & $   0.923  \pm 0.020  $&  $   0.9348 \pm 0.0001  $&  $ -0.6     $&  $ -0.6     $ \cr\noalign{\vskip 1pt}
$A_c$                            & $   0.670  \pm 0.027  $&  $   0.6679 \pm 0.0005  $&  $  0.1     $&  $  0.1     $ \cr\noalign{\vskip 1pt}
$A_s$                            & $   0.895  \pm 0.091  $&  $   0.9357 \pm 0.0001  $&  $ -0.4     $&  $ -0.4     $ \cr\noalign{\vskip 1pt}
\end{tabular}}
\label{ch7_zpole}
\end{table}

\begin{table}[htdp]
\tbl{Principal non-$Z$-pole observables,  as of 11/07, from~\refcite{Amsler:2008zz}.
$m_t$ is from the direct CDF and D0 measurements at the Tevatron;
$M_W$ is determined mainly by CDF, D0, and the LEP II collaborations;
$g_{L}^2,$ corrected for the $s-\bar s$ asymmetry, and $g_R^2$ are from NuTeV; 
$g_V^{\nu e}$  are dominated by the CHARM II experiment at CERN;
$A_{PV}$ is from the SLAC polarized  M\o ller asymmetry; and the $Q_W$ are from atomic parity violation. 
}
{\begin{tabular}{lllrr} 
 Quantity & {\rm Value} & {\rm Standard Model} & {\rm Pull} & {\rm Dev.}                       \cr 
\hline \noalign{\vskip 1pt}
$m_t$                    [GeV]  &$ 170.9\pm 1.8\pm 0.6 $& $171.1    \pm 1.9$     &$ -0.1$     &$ -0.8  $   \cr\noalign{\vskip 1pt}
$M_W$   ($\bar p p$)   & $ 80.428  \pm 0.039$  & $ 80.375  \pm 0.015 $  &  1.4     &  1.7     \cr\noalign{\vskip 1pt}
$M_W$ (LEP)    \quad                            & $ 80.376  \pm 0.033 $ &                      &  0.0     &  0.5     \cr\noalign{\vskip 1pt}
$g_L^2$                          &  $ 0.3010 \pm 0.0015$ &$   0.30386\pm 0.00018 $&$ -1.9    $ &$ -1.8 $    \cr\noalign{\vskip 1pt}
$g_R^2$                          &   $0.0308 \pm 0.0011$ & $  0.03001\pm 0.00003$ &  0.7     &  0.7     \cr\noalign{\vskip 1pt}
$g_V^{\nu e}$                    & $ -0.040  \pm 0.015 $ & $ -0.0397 \pm 0.0003 $ &  0.0     &  0.0      \cr\noalign{\vskip 1pt}
$g_A^{\nu e}$                    & $ -0.507  \pm 0.014 $ & $ -0.5064 \pm 0.0001 $ &  0.0     &  0.0     \cr\noalign{\vskip 1pt}
$A_{PV} \times 10^7$ &$ -1.31 \pm 0.17$ &$ -1.54 \pm 0.02 $ &  1.3     &  1.2      \cr\noalign{\vskip 1pt}
$Q_W({\rm Cs})$                  &$ -72.62   \pm 0.46 $  &$ -73.16   \pm 0.03$    &  1.2     &  1.2     \cr\noalign{\vskip 1pt}
$Q_W({\rm Tl})$                  &$-116.4    \pm 3.6 $   &$-116.76   \pm 0.04    $&  0.1     &  0.1     \cr\noalign{\vskip 1pt}
\end{tabular}}
\label{ch7_nonzpole}
\end{table}

The data are sensitive to $m_t$, $\alpha_s$ (evaluated at $M_Z$), and $M_H$, which enter the radiative
corrections. The precision data alone yield 
$m_t = 174.7^{+10.0}_{-7.8}$ GeV, in impressive
agreement with the direct Tevatron value $170.9 \pm 1.9$. The $Z$-pole data alone yield
 $\alpha_s=0.1198(20)$, in good agreement with the world average of $0.1176(20)$,
 which includes other determinations at lower scales. The higher value in \refl{results}
 is due to the inclusion of data from hadronic $\tau$ decays\footnote{A recent reevaluation of the
 theoretical formula~\cite{Maltman:2008ud} lowers the $\tau$ value to $0.1187(16)$, consistent with the other determinations.}. 
 
\begin{figure}[htbp]
\centering
\includegraphics*[scale=0.45]{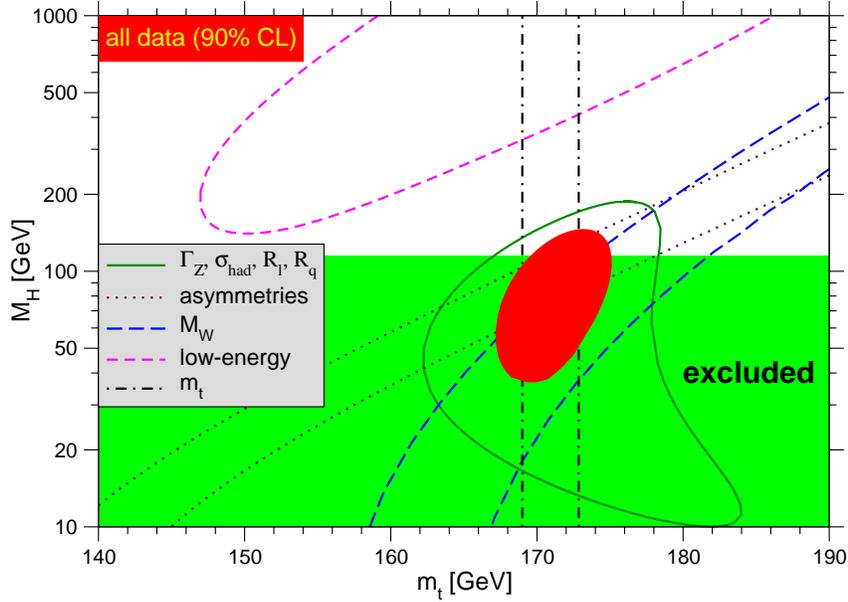}
\caption{1$\sigma$ allowed regions in \mh \  vs $m_t$ \ and the 90\% cl global fit region from precision data,
compared with the direct exclusion limits from LEP 2. Plot courtesy of the Particle Data Group~\cite{Amsler:2008zz}.}
\label{higgspdf}
\end{figure}

 The prediction for the Higgs mass from indirect data\footnote{The predicted value would decrease if new physics accounted for
the value of $A_{FB}^{(0b)}$~\cite{Chanowitz:2002cd}.}, 
\mh $= 77^{+28}_{-22}$ GeV, should be compared with the 
direct LEP 2 limit 
$\mh \gtrsim 114.4\, (95\%)$ GeV~\cite{Barate:2003sz}. There is no direct conflict given the large
uncertainty in the prediction, but the central value is in the excluded region,
as can be seen in Figure \ref{higgspdf}.
 Including the direct LEP 2 exclusion results, one finds
$\mh < 167$ GeV at 95\%. As of this writing CDF and D0 are becoming sensitive to the upper end of this
range, and have a good chance of discovering or excluding the SM Higgs in the entire allowed region.
We saw in Section \ref{higgssection}
that there is a theoretical range  $115 \text{ GeV } < M_H < 180 \text{ GeV } $ in the SM
provided it is valid up to the Planck scale, with a much wider allowed range otherwise.
The experimental constraints on \mh\
  are encouraging for  supersymmetric extensions of the SM, which involve more complicated Higgs sectors.
The quartic Higgs self-interaction $\lambda$ in \refl{eq15} is replaced by gauge couplings, 
leading to a theoretical  upper limit  $\mh \lesssim 130$ GeV
in the minimal supersymmetric extension (MSSM), while \mh \ can be as high as 150 GeV in generalizations.
 In the decoupling limit
 in which the second Higgs doublet is much heavier
 the direct search lower limit is similar to the standard model. However, the
 direct limit is
  considerably lower in the non-decoupling region in which the new supersymmetric particles and second Higgs 
  are relatively light~\cite{Heinemeyer:2004gx,Heinemeyer:2007bw,Barate:2003sz}. 

\begin{figure}[htbp]
\begin{center}
\includefigure{0.32}{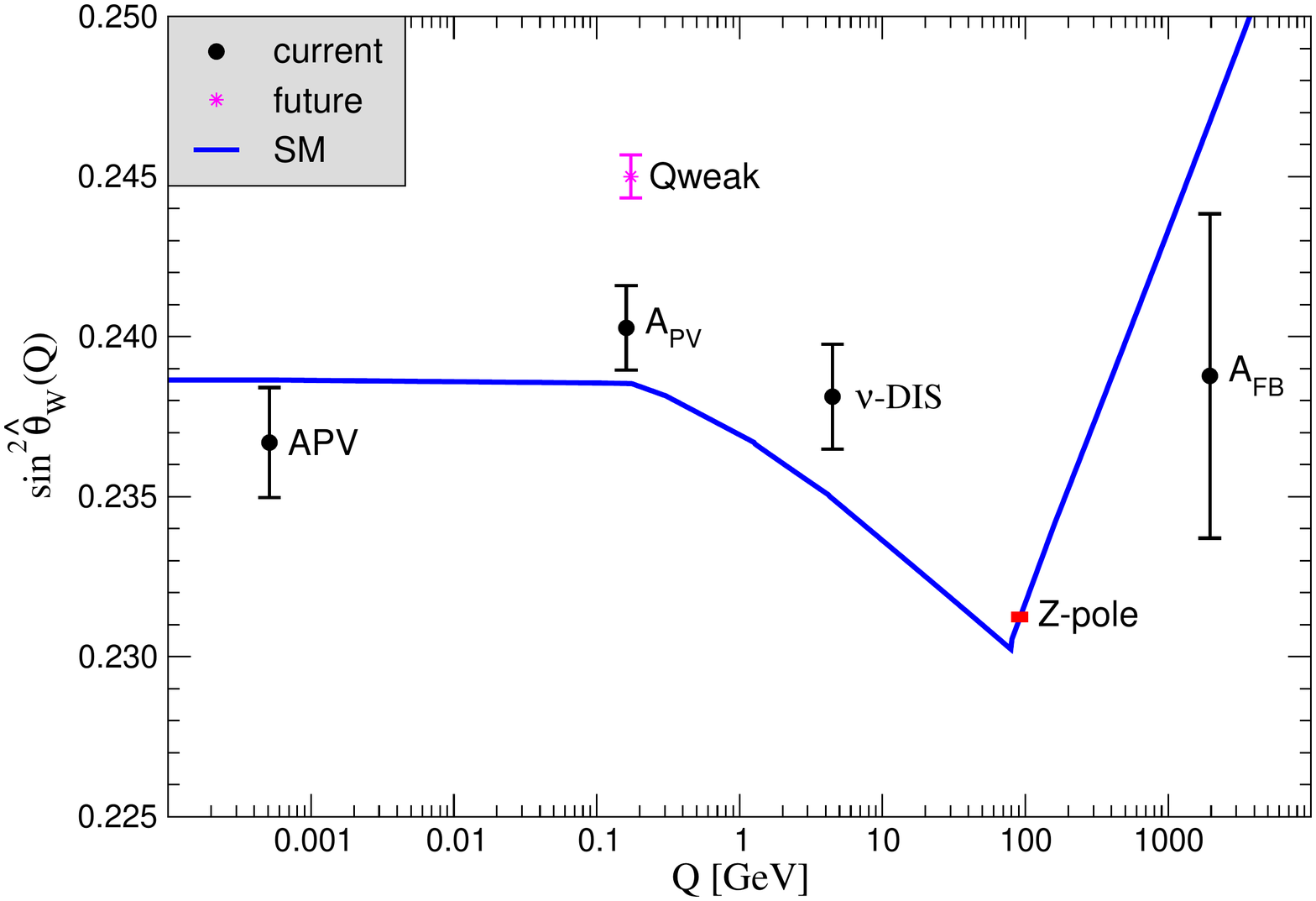}
\end{center}
\caption{Running $\hat s_Z^2 (Q^2)$ measured at various scales, compared with the 
predictions of the SM~\cite{Czarnecki:2000ic}. The low energy points
are from atomic parity violation (APV), the polarized  M\o ller asymmetry (PV)
and deep inelastic neutrino scattering (corrected for an $s-\bar s$ asymmetry).
$Q_{weak}$ shows the expected sensitivity of a future polarized $e^-$ measurement
at Jefferson Lab.
Plot courtesy of the Particle Data Group~\cite{Amsler:2008zz}.}
\label{s2w_2007}
\end{figure}

It is interesting to compare the $Z$ boson couplings measured at different energy scales. 
The renormalized weak angle measured at different scales in the \msb\ scheme is displayed in
Figure \ref{s2w_2007}.

The precision program has also been used to search for and constrain the effects of possible new TeV scale physics\footnote{For reviews, see \refcite{Langacker:1991zr,Erler:2003yk,Erler:2008ek,Amsler:2008zz}.}.  This includes the effects of possible mixing between ordinary and exotic heavy fermions~\cite{Langacker:1988ur}, new $W'$ or $Z'$ gauge bosons~\cite{Hewett:1988xc,Langacker:2008yv},
leptoquarks~\cite{Cheung:2001wx,Chemtob:2004xr,Barbier:2004ez,Alcaraz:2006mx}, Kaluza-Klein excitations in extra-dimensional theories~\cite{Hewett:2002hv,Csaki:2004ay,Sundrum:2005jf,Amsler:2008zz}, and new four-fermion operators~\cite{Cho:1997kf,Cheung:2001wx,Han:2004az,Alcaraz:2006mx},
all of which can effect the observables at tree level. The oblique corrections~\cite{Peskin:1990zt,Peskin:1991sw}, which only affect the
$W$ and $Z$ self energies, are also constrained. The latter may be generated, e.g., by heavy non-degenerate scalar or fermion multiplets and heavy
chiral fermions~\cite{Amsler:2008zz}, such as are often found in models that replace the elementary Higgs by a dynamical mechanism~\cite{Hill:2002ap}.
A major implication of supersymmetry is through the small mass expected for the lightest Higgs boson.  Other
supersymmetric effects are small in the decoupling limit in which the superpartners and extra Higgs doublet are heavier than a few hundred GeV~\cite{Erler:1998ur,Heinemeyer:2004gx,Heinemeyer:2007bw,Ellis:2007fu}.
The precisely measured gauge couplings at the $Z$-pole are also important for 
testing the ideas of gauge coupling unification~\cite{Georgi:1974yf}, which works extremely well in the MSSM~\cite{Amaldi:1991cn,Ellis:1990zq,Giunti:1991ta,Langacker:1991an}. 

\subsection{Gauge Self-interactions}
The \st\ gauge kinetic energy terms in \refl{eqch11b} lead to 3 and 4-point gauge self-interactions for
the $W$'s,
\beq  
\begin{split}
\lag_{W3} = &-ig \bigl(\partial_\rho W^3_\nu \bigr) W^+_\mu W^-_\sigma \bigl[ g^{\rho\mu} g^{\nu\sigma}
   -g^{\rho\sigma}g^{\nu\mu}\bigr] \\
&   -ig \bigl(\partial_\rho  W^+_\mu \bigr) W^3_\nu W^-_\sigma \bigl[ g^{\rho\sigma} g^{\mu\nu}
   -g^{\rho\nu}g^{\mu\sigma}\bigr] \\
  &   -ig \bigl(\partial_\rho  W^-_\sigma \bigr) W^3_\nu W^+_\mu \bigl[ g^{\rho\nu} g^{\mu\sigma}
   -g^{\rho\mu}g^{\nu\sigma}\bigr],
\end{split}\eeql{smlag1}
and
\beq  
\lag_{W4} = \frac{g^2}{4} \left[ W^+_\mu W^+_\nu W^-_\sigma W^-_\rho {\cal Q}^{\mu\nu\rho\sigma} 
-2 W^+_\mu W^3_\nu W^3_ \sigma W^-_ \rho {\cal Q}^{\mu\rho\nu\sigma}\right],
\eeql{smlag2}
where 
\beq  {\cal Q}_{\mu\nu\rho\sigma} \equiv 2 g_{\mu\nu}g_{\rho\sigma}- g_{\mu \rho}g_{\nu\sigma}
- g_{\mu \sigma}g_{\nu \rho}.
\eeql{smlag3}
These  carry over to the $W$, $Z$, and $\gamma$ self-interactions
provided we replace $W^3$ by $\cos \theta_W Z + \sin \theta_W A$ using \refl{eqch33} and
\refl{eqch34} (the $B$ has no self-interactions). The resulting vertices
follow from the matrix element of $i\lag$ after including identical particle factors
and using $g=e/\sin \theta_W$. They
 are listed in Table \ref{ch7_smrules} and shown in Figure \ref{figure16}.
\newlength{\wself}
\setlength{\wself}{0.8cm}
\begin{figure}
\begin{minipage}[t]{5.0cm} 
\includegraphics*[scale=0.8]{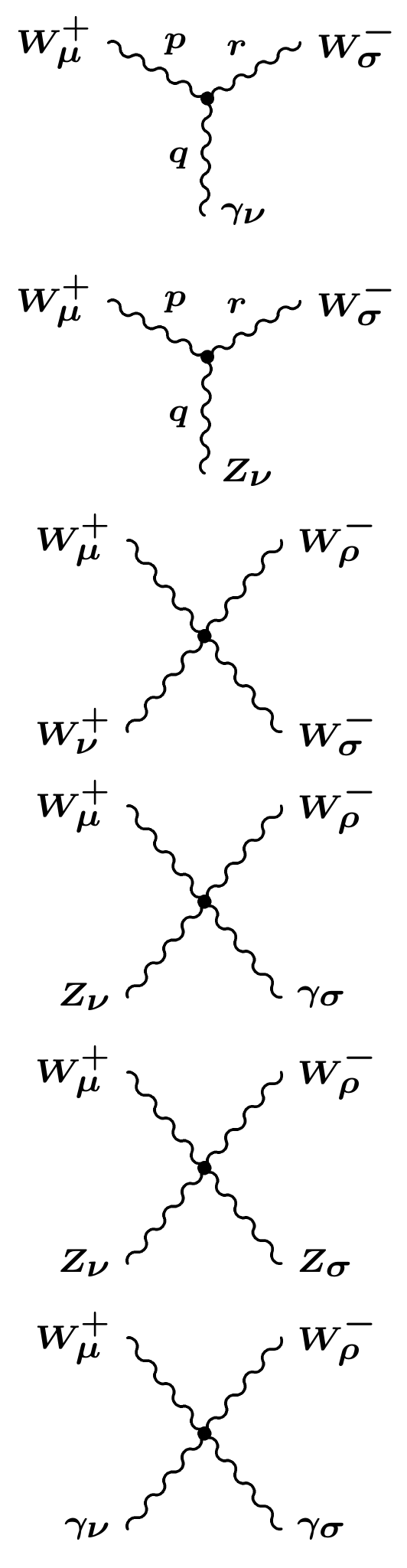}
\end{minipage}
\hspace*{1.2cm}
\begin{minipage}[b]{4cm} \normalsize
\[  ie {\cal C}_{\mu\nu\sigma}(p,q,r) \]
\[ \begin{split}
&{\cal C}_{\mu\nu\sigma}(p,q,r)\equiv  g_{\mu \nu} ( q - p )_{\sigma} 
\\ & \quad +g_{\mu \sigma} (p - r)_{\nu} + g_{\nu \sigma} (r - q)_{\mu} 
\end{split} \]
\vspace*{0.8\wself}
\[  i\frac{e}{\tan \theta_W}  {\cal C}_{\mu\nu\sigma}(p,q,r) \]
\vspace*{1.7\wself}
\[ i\frac{e^2}{\sin^2 \theta_W}  {\cal Q}_{\mu\nu\rho\sigma}  \]
\[\begin{split}
{\cal Q}_{\mu\nu\rho\sigma} \equiv& 2 g_{\mu\nu}g_{\rho\sigma}\\ &- g_{\mu \rho}g_{\nu\sigma}
- g_{\mu \sigma}g_{\nu \rho} 
\end{split}\]
\vspace*{0.4\wself}
\[ - i\frac{e^2}{\tan \theta_W}  {\cal Q}_{\mu\rho\nu\sigma} \]
\vspace*{1.7\wself}
\[ - i\frac{e^2}{\tan^2 \theta_W}  {\cal Q}_{\mu\rho\nu\sigma} \] 
\vspace*{1.7\wself}
\[ - ie^2 {\cal Q}_{\mu\rho\nu\sigma} \]
\vspace*{1\wself}
\end{minipage}
\caption{The three and four point-self-interactions of gauge bosons in
the standard electroweak model. The momenta and charges flow into the vertices.}
\label{figure16}
\end{figure}

 The gauge self-interactions are essential probes of the structure and consistency of a spontaneously-broken
non-abelian gauge theory. Even tiny deviations in their form or value would
destroy the delicate cancellations needed for renormalizability,
and would signal the need either for compensating new physics (e.g., from mixing
with other gauge bosons or new particles in loops), or of a more fundamental
breakdown of the gauge principle, e.g., from some forms of compositeness.
They have been constrained by measuring the total cross section and various decay distributions for
 $e^-e^+\ra W^-W^+$ at LEP~2, and by observing $\bar p p\ra W^+ W^-, WZ,$ and $W\gamma$ at the
 Tevatron. Possible anomalies in the predicted quartic vertices in Table  \ref{ch7_smrules},
 and the neutral cubic vertices for  $ZZZ,$ $ZZ\gamma,$ and $ Z\gamma\gamma$, which are absent in the SM,
 have also been constrained by LEP~2~\cite{Alcaraz:2006mx}.
 
\begin{figure}
\centering
\includegraphics*[scale=0.55]{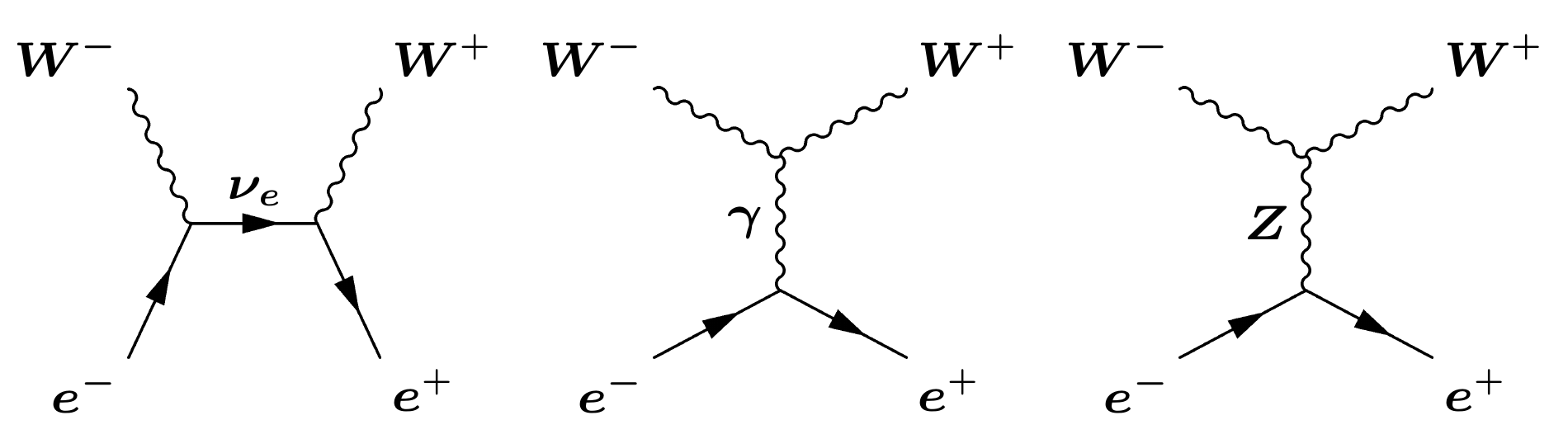}
\caption{Tree-level diagrams contributing to $e^+e^- \ra
W^+W^-$.}
\label{figure3}
\end{figure}

\begin{figure}[htbp]
\begin{center}
\includefigure{.4}{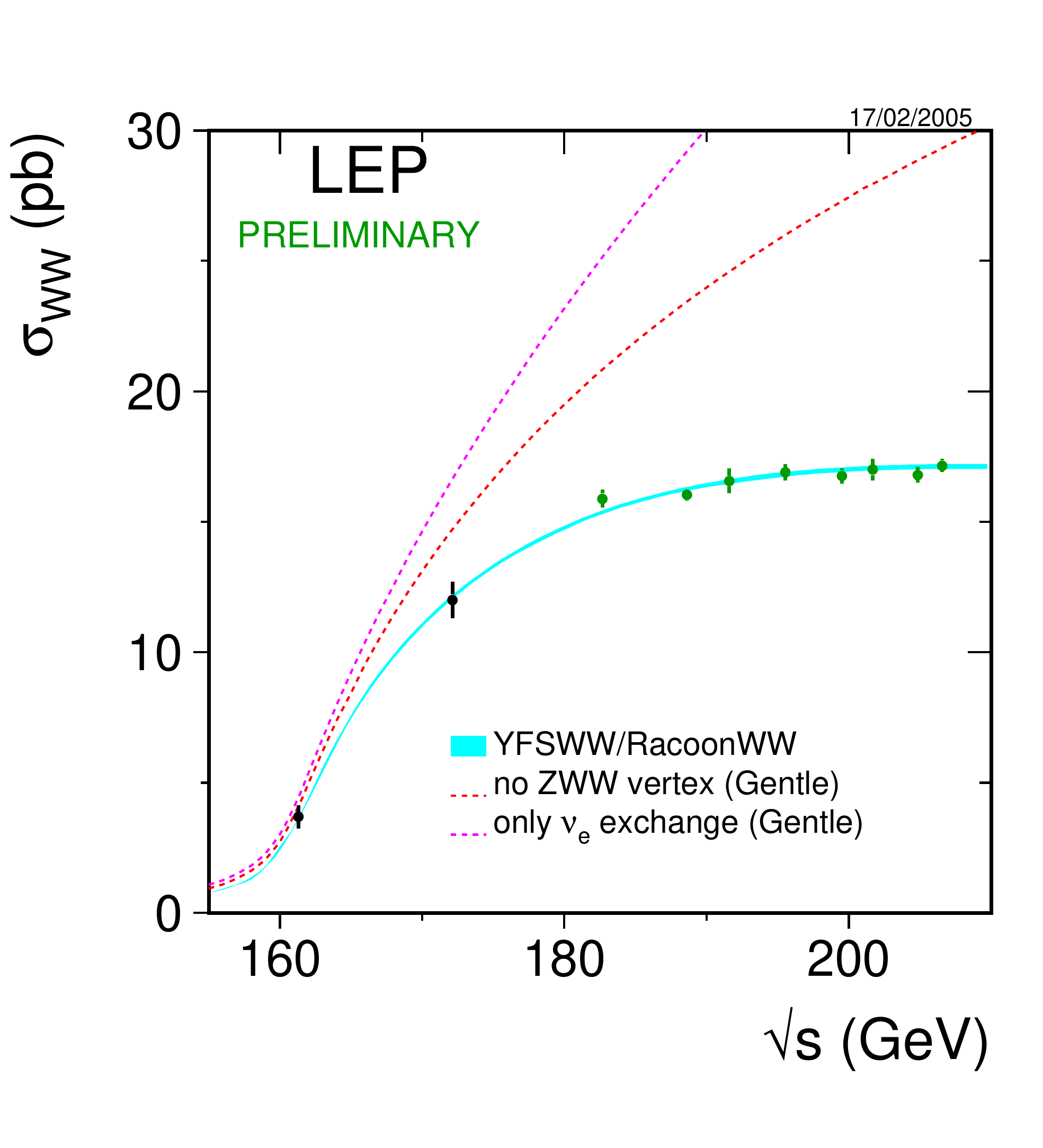}
\end{center}
\caption{Cross section for $e^-e^+\ra W^-W^+$ compared with the SM expectation. Also
shown is the expectation from $t$ channel $\nu_e$ exchange only, and
for the $\nu_e$ and $\gamma$ diagrams only.  Plot courtesy of the LEP Electroweak Working Group~\cite{Alcaraz:2006mx},
{\tt http://www.cern.ch/LEPEWWG/}.}
\label{eeww}
\end{figure}

The three tree-level diagrams for $e^-e^+\ra W^-W^+$ are shown in Figure \ref{figure3}.
The cross section from any one or two of these rises rapidly with center of mass energy, but gauge 
invariance relates these
three-point vertices to the couplings of the fermions in such a way that at
high energies there is a cancellation.  It is another manifestation of the
cancellation in a gauge theory which brings higher-order loop integrals under control,
leading to a renormalizable theory. It is seen in Figure \ref{eeww} that the expected cancellations do occur.

\section{Problems with the Standard Model}\label{problemssm}

For convenience we summarize the  Lagrangian density 
after spontaneous symmetry breaking:
\beq \begin{split}
\lag &= \lag_{gauge} + \lag_\vp
   + \sum_r \bar{\psi}_r \left( i \not{\!\partial}- m_r - \frac{m_r
H}{\nu} \right) \psi_r \\
&- \frac{g}{2 \sqrt{2}} \left( J^\mu_W W^-_\mu + J^{\mu \dag}_W
W^+_\mu \right)
   - e J^\mu_{Q} A_\mu - \frac{g}{2\cos \theta_W} J^\mu_Z Z_\mu,
\end{split}
\eeql{ch8:pr0}
where the self-interactions for the $W^\pm$, $Z$, 
and $\gamma$ are given in \refl{smlag1} and \refl{smlag2},  $\lag_\vp$ is given in \refl{eq7:ssb18},
and the fermion currents in \refl{eqch50}, \refl{eqch57}, and \refl{eqch60}.
For Majorana $\nu_L$ masses generated by a higher dimensional operator involving two factors of the Higgs doublet, as in the
seesaw model, the $\nu$ term in \refl{ch8:pr0} is replaced by
\beq  
\lag=\sum_r \bar \nu_{rL} i  \not{\!\partial} \nu_{rL} -\oh m_{\nu r} \Bigl( \bar \nu_{rL} \nu^c_{rR} + h.c. \Bigr) \Bigl(1+\frac{H}{\nu} \Bigr)^2,
\eeql{smlag4}
where $\nu^c_{rR}$ is the $CP$ conjugate to $\nu_L$ (see, e.g.,~\refcite{Langacker:2005pfa}).

The standard electroweak model is a mathematically-consistent
renormalizable field theory which predicts or is consistent with all
experimental facts.  It successfully predicted the existence and form
of the weak neutral current, the existence and masses of the $W$ and
$Z$ bosons, and the charm quark, as necessitated by the GIM mechanism.
The charged current weak interactions, as described by the generalized
Fermi theory, were successfully incorporated, as was quantum
electrodynamics. 
The consistency between theory and experiment indirectly tested the radiative corrections and
ideas of renormalization and allowed the successful prediction of the top quark mass.
 Although the original formulation did not provide for massive neutrinos,
they are easily incorporated by the addition of right-handed states $\nu_R$ (Dirac)
or as higher-dimensional operators, perhaps generated by an underlying seesaw (Majorana).
When combined with quantum chromodynamics for the
strong interactions, the
standard model is almost certainly the approximately correct
description of the elementary particles and their interactions down to at least $10^{-16}$cm, with the
possible exception of the Higgs sector or new very weakly coupled particles.  When combined with 
general relativity for classical gravity the SM accounts for most of the observed features of Nature
(though not for the dark matter and energy).

However, the
theory has far too much arbitrariness to be the final story.  For
example, the minimal version of the model has 20 free parameters for massless neutrinos 
and another 7 (9) for massive  Dirac (Majorana) neutrinos\footnote{12 fermion masses (including the neutrinos), 6 mixing angles, 2 $CP$ violation phases 
($+$ 2 possible Majorana phases), 3 gauge couplings, $M_H$, $\nu$, $\theta_{QCD}$, $M_P$, $\Lambda_{cosm}$, minus one overall mass scale since only mass ratios
are physical.},
not counting electric charge (i.e., hypercharge)
assignments. 
 Most physicists believe that this is just too much for
the fundamental theory.  The complications of the standard model can
also be described in terms of a number of problems.

\subsection*{The Gauge Problem}
The standard model is a complicated direct product of three
subgroups, \sthto,   
with separate gauge couplings.
There is no explanation for why only the electroweak part is chiral
(parity-violating).  Similarly, the standard model incorporates but
does not explain another fundamental fact of nature: {charge
quantization}, 
{i.e.,} why all particles have charges which are
multiples of $e/3$.  This is important because it allows the
electrical neutrality of atoms $(|q_p| = |q_e|)$.  
The complicated gauge structure suggests the existence of some underlying
unification of the interactions, such as one would expect in a superstring~\cite{Green:110204,Polchinski:363850,Becker:1003112} or grand unified theory~\cite{Georgi:1974sy,Langacker:1980js,Ross:1107830,Hewett:1988xc,Raby:2008gh}.
Charge quantization can also be explained in such theories,
though the ``wrong'' values of charge emerge in some constructions due to different hypercharge embeddings
or non-canonical values of $Y$ (e.g., some string constructions lead to exotic particles with
charges of $\pm e/2$).
Charge quantization may also be explained, at least in part, by the existence of
magnetic monopoles~\cite{Preskill:1984gd} or the absence of anomalies\footnote{The absence of anomalies
is not sufficient to determine all of the $Y$ assignments without
additional assumptions, such as family universality.}, but either of these is likely to find its origin
in some kind of underlying unification.

\subsection*{The Fermion Problem}

All matter under ordinary terrestrial conditions can be constructed
out of the fermions $ (\nu_e, e^-, u, d)$ of the first family.  Yet we
know from laboratory studies that there are $\geq 3$ families:
$(\nu_\mu, \mu^-, c, s)$ and $(\nu_\tau, \tau^-, t, b) $ are heavier
copies of the first family with no obvious role in nature.  
The standard model gives no
explanation for the existence of these heavier families and no
prediction for their numbers.  Furthermore, there is no explanation or
prediction of the fermion masses, 
which are observed to occur in a hierarchical pattern which varies over 5 orders of magnitude
between the $t$ quark and the $e^-$, or of the quark and lepton mixings.  
Even more mysterious are the neutrinos, which are many orders of magnitude lighter still.
It is not even certain whether the neutrino masses are Majorana or Dirac.
A related difficulty is that while the $CP$ violation observed in the laboratory is
well accounted for by the phase in the CKM matrix, there is no SM source of $CP$ breaking
adequate to explain the baryon asymmetry of the universe.

There are many possible
suggestions of new physics that might shed light on these questions.
The existence of multiple families could be due to large representations of
some string theory or grand unification, or they could be associated with different
possibilities for localizing particles in some higher dimensional space. The latter could also be associated 
with string compactifications, or by some effective brane world scenario~\cite{Hewett:2002hv,Csaki:2004ay,Sundrum:2005jf,Amsler:2008zz}. The hierarchies of masses
and mixings could emerge from wave function overlap effects in such higher-dimensional spaces. Another interpretation, also
possible in string theories, is that the hierarchies are because some of the mass terms
are generated by higher dimensional operators and therefore suppressed by powers of
$\vev{S}/M_X$, where $S$ is some standard model singlet field and $M_X$ is some large scale such as $M_P$.  The allowed operators could perhaps be enforced by some family 
symmetry~\cite{Froggatt:1978nt}. Radiative hierarchies~\cite{Babu:1990vx}, in which some of the masses are generated at the loop level, or some form of compositeness are other possibilities. 
Despite all of these ideas there is no
compelling model and none of these yields detailed predictions.  
  Grand unification by itself doesn't
help very much, except for the prediction of $m_b$ in terms
of $m_\tau$ in the simplest versions.  

The small values for the neutrino masses suggest
that they are associated with Planck or grand unification physics, as in the seesaw model,
but there are other possibilities~\cite{GonzalezGarcia:2002dz,Langacker:2005pfa,Mohapatra:2005wg,GonzalezGarcia:2007ib}.

Almost any type of new physics  is likely to lead to new sources of $CP$ violation.

\subsection*{The Higgs/Hierarchy Problem}
In the standard model one introduces an elementary Higgs field 
 to generate masses for the $W$, $Z$, and
fermions. For the model to be consistent the
Higgs mass should not be too different from the $W$ mass.  If $M_H$ were to be  larger than $M_W$ by many orders of
magnitude the Higgs
self-interactions would be excessively strong.  Theoretical
arguments suggest that  $M_H \lesssim 700$~GeV
(see Section \ref{higgssection}).
 
\begin{figure}[htbp]
\begin{center}
\includefigure{0.6}{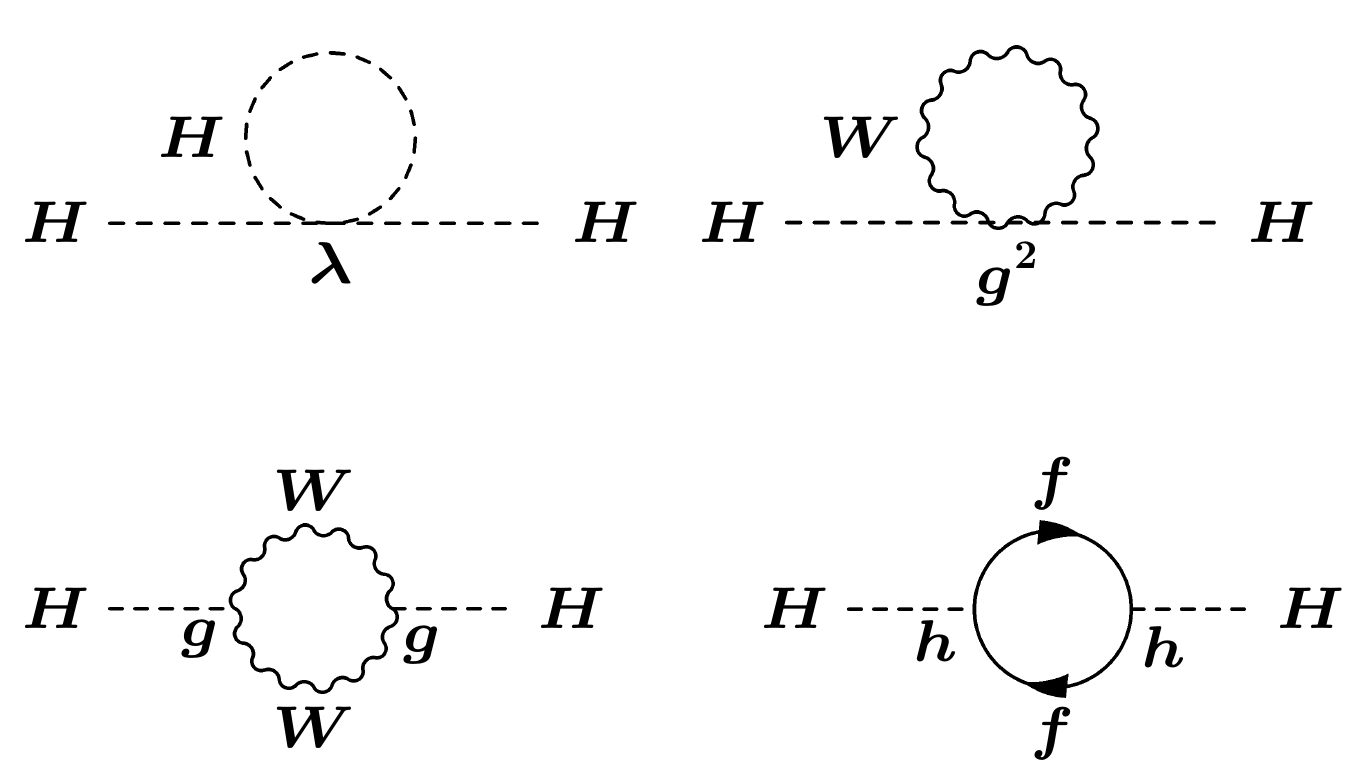}
\end{center}
\caption{Radiative corrections to the Higgs mass, including
self-interactions, interactions with gauge bosons, and interactions with
fermions.}
\label{Higgsmass}
\end{figure}

However, there is a complication.  The tree-level (bare) Higgs mass
receives quadratically-divergent corrections from the loop diagrams in
Figure~\ref{Higgsmass}.
One finds
\beq M_H^2 = (M_H^2)_{bare} + {\cal O} (\lambda, g^2, h^2) \Lambda^2,
\eeql{ch8:pr1}
where $\Lambda$ is the next higher scale in the theory.  If there were
no higher scale one could simply interpret $\Lambda$ as an ultraviolet
cutoff and take the view that $M_H$ is a measured parameter, with
$(M_H)_{bare}$ not observable.  However, 
 the theory is
presumably embedded in some larger theory that cuts off the momentum integral
at the finite scale of the new physics\footnote{There is no analogous
fine-tuning associated with logarithmic divergences, such as those
encountered in QED, because $\alpha \ln (\Lambda/m_e) < {\cal O}(1)$ even
for $\Lambda = M_P$.}.  
For example, if the next
scale is gravity $\Lambda$ is the Planck scale $M_P = G_N^{-1/2} \sim
10^{19}$~GeV.  In a grand unified theory, 
one would expect $\Lambda$ to be of order the unification scale $M_X
\sim 10^{14}$~GeV.  Hence, the natural scale for $M_H$ is
${\cal O}(\Lambda)$, which is much larger than the expected value.  There
must be a fine-tuned and apparently highly contrived
cancellation between the bare value and the correction, to more than
30 decimal places in the case of gravity.  If the cutoff is provided
by a grand unified theory there is a separate hierarchy problem at the
tree-level.  The tree-level couplings between the Higgs field and the
superheavy fields lead to the expectation
that $M_H$ is close to the unification scale unless unnatural
fine-tunings are done, i.e., one does not understand why $(M_W/M_X)^2$ is so
small in the first  place.

One solution to  this Higgs/hierarchy problem is TeV scale supersymmetry,
in which the quadratically-divergent contributions of fermion and boson loops
cancel, leaving only much smaller effects of the order of supersymmetry-breaking.
(However, supersymmetric grand unified theories still suffer from the tree-level hierarchy problem.)
There are also (non-supersymmetric) extended models in which the cancellations 
are between bosons or between fermions. This class includes Little Higgs models~\cite{ArkaniHamed:2002qy,Perelstein:2005ka}, in which the Higgs is forced to be lighter than new TeV scale
dynamics because it is a pseudo-Goldstone boson of an approximate underlying global symmetry,
and Twin-Higgs models~\cite{Chacko:2005pe}.

Another possibility is to eliminate the elementary Higgs fields, replacing them  with some {dynamical symmetry breaking} mechanism based on a new strong dynamics~\cite{Hill:2002ap}. In technicolor, for example,  the SSB is associated with the expectation
value of a fermion bilinear, analogous to the breaking of chiral symmetry in QCD. Extended technicolor,
top-color, and composite Higgs models all fall into this class.

Large and/or warped extra dimensions~\cite{ArkaniHamed:1998rs,Dienes:1998vg,Randall:1999ee} can also resolve the difficulties, by altering the relation
between $M_P$ and a much lower fundamental scale, by providing a cutoff at the inverse of the extra dimension scale, or by using the boundary conditions in the extra
dimensions to break the electroweak symmetry (Higgsless models~\cite{Csaki:2003zu}).
Deconstruction models, in which no extra dimensions are explicity introduced~\cite{ArkaniHamed:2001nc,Hill:2000mu}, are closely related.

Most of the models mentioned above have the potential to generate flavor changing neutral current
and $CP$ violation effects much larger than observational limits.  Pushing the mass scales high enough to avoid these problems may  conflict with a natural
solution to  the hierarchy problem, i.e., one
may reintroduce a  {little hierarchy problem}.  Many  are also strongly constrained  by precision electroweak physics. In some cases the new physics does not satisfy the decoupling theorem~\cite{Appelquist:1974tg}, leading to large oblique corrections. In others new tree-level effects may again force the scale to be  too high. The most successful from the precision electroweak
point of view are those which have a discrete symmetry which prevents vertices involving just one
heavy particle, such as $R$-parity in supersymmetry, $T$-parity in some little Higgs models~\cite{Cheng:2003ju},
and $KK$-parity in universal extra dimension models~\cite{Appelquist:2000nn}.

A very different possibility is to accept the fine-tuning, i.e.,  to abandon the notion of {naturalness} for the
weak scale,
perhaps motivated by {anthropic}
considerations~\cite{Agrawal:1997gf}.  (The anthropic idea will be considered below in the discussion of the gravity problem.) This could emerge, for example, in split supersymmetry~\cite{ArkaniHamed:2004fb}.

\subsection*{The Strong $CP$ Problem}
Another fine-tuning problem is the {strong $CP$ problem}~\cite{Peccei:1988ci,Dine:2000cj,Kim:2008hd}.
One can add an additional term $\frac{\theta_{QCD}}{32 \pi^2} g^2_s G
\tilde{G}$ to the QCD Lagrangian density which breaks $P$, $T$ and $CP$ symmetry\footnote{One could add an analogous term
for the weak \st\ group, but it does not lead to observable consequences, at least within
the SM~\cite{Anselm:1993uj,Dine:2000cj}.}.
$\tilde{G}_{\mu \nu}^i = \epsilon_{\mu \nu \alpha \beta} G^{\alpha
\beta i}/2$ is the dual field strength  tensor.
This term, if present, would induce an electric dipole moment $d_N$ for
the neutron.  The rather stringent limits on the dipole
moment
lead to the upper bound $|\theta_{QCD}| <10^{-11}$.             
  The question is, therefore, why is $\theta_{QCD}$
so small?  It is not sufficient to just say that it is zero (i.e., to impose
$CP$ invariance on QCD) because
of the observed violation of $CP$ by the weak interactions.
As discussed in Section \ref{wccints}, this is believed to be associated with phases in the
quark mass matrices. The quark phase redefinitions which remove them lead to a shift in 
 $\theta_{QCD}$ by $\mathcal{O} (10^{-3})$ because of the anomaly in the vertex coupling the
 associated global current to two gluons.
Therefore, an apparently contrived fine-tuning is needed to cancel
this correction against the bare value.  Solutions include the
possibility that $CP$ violation is not induced directly by phases in
the Yukawa couplings, as is usually assumed in the standard model, but
is somehow violated spontaneously.
 $\theta_{QCD}$  then
would be a calculable parameter induced at loop level, and it is
possible to make $\theta_{QCD}$ sufficiently small.  However, such models
lead to difficult phenomenological and cosmological problems\footnote{Models
in which the $CP$ breaking occurs near the Planck scale may
be viable~\cite{Nelson:1984hg,Barr:1984fh}.}.
Alternately, $\theta_{QCD}$ becomes unobservable
(i.e., can be rotated away)
if there is a massless $u$ quark~\cite{Kaplan:1986ru}.
However, most  phenomenological estimates~\cite{Nelson:2003tb} are not consistent with
$m_u = 0$.
Another possibility is the Peccei-Quinn mechanism~\cite{Peccei:1977hh},
in which an extra global $U(1)$ symmetry is imposed on the theory in such a way that
$\theta_{QCD}$ becomes a dynamical variable which is zero at the minimum of
the potential.  
The spontaneous breaking of the symmetry, along with 
explicit breaking associated with
the anomaly and instanton effects, leads to a very
light pseudo-Goldstone boson known as an axion~\cite{Weinberg:1977ma,Wilczek:1977pj}.
Laboratory, astrophysical, and
cosmological constraints suggest the range $10^9 - 10^{12}$~GeV for
the scale at which the $U(1)$ symmetry is broken.

\subsection*{The Gravity Problem}\label{gravity}
Gravity is not fundamentally unified with the other interactions in
the standard model, although it is possible to graft on classical
general relativity by hand.  However, general relativity is not a quantum theory,
and there is no obvious way to generate one within the
standard model context.   Possible
solutions include Kaluza-Klein~\cite{Chodos:109424}
and supergravity~\cite{Nilles:1983ge,Wess:320631,Terning:941399} theories.
These connect gravity with the other interactions in a more natural way, but do not 
yield renormalizable theories of quantum
gravity. More promising are superstring theories (which may incorporate
the above), which unify gravity
and may  yield {\em finite} theories of quantum gravity and all the
other interactions. String theories are perhaps the most likely possibility
for the underlying theory of particle physics and gravity, but at present there 
appear to be a nearly unlimited number of possible string vacua (the landscape),
with no obvious selection principle. As of this writing the particle physics community
is still trying to come to grips with the landscape and its implications. Superstring theories naturally
imply some form of supersymmetry, but it could be broken at a high scale
and have nothing to do with the Higgs/hierarachy problem (split supersymmetry
is a compromise, keeping some aspects at the TeV scale).

In addition to the fact that gravity is not
unified and not quantized there is another difficulty, namely the
cosmological constant.  The cosmological constant can be thought of as the
energy of the vacuum. However,
we saw in Section \ref{higgssection} that the spontaneous breaking of \sto\
generates  a value
$\vev{V(\nu)}=-\mu^4/4\lambda$ for the expectation value of the Higgs
potential at the minimum.
 This is a $c$-number
which has no significance for the microscopic interactions.
However, it assumes great importance when the theory is coupled to
gravity, because it contributes to the
cosmological constant.
The cosmological
constant becomes
\beq \Lambda_{cosm} = \Lambda_{bare} + \Lambda_{SSB} ,
\eeql{ch8:pr2}
where $\Lambda_{bare}=8\pi G_N V(0)$ is the primordial cosmological constant,
which can be thought of as the value of the energy of the vacuum in
the absence of spontaneous symmetry breaking. 
(The definition of $V(\vp)$ in \refl{eq15}
implicitly assumed $\Lambda_{bare} = 0$.) $\Lambda_{SSB}$ is
the part generated by the Higgs mechanism:
\beq  |\Lambda_{SSB}| = 8 \pi G_N |\vev{V}|
\sim 10^{56} \Lambda_{obs} .
\eeql{ch8:pr3}
It is some $10^{56}$ times larger in magnitude than the observed value
$\Lambda_{obs}\sim(0.0024 \text{ eV})^4/8 \pi G_N$ (assuming that the dark energy is due to a cosmological constant),
and it is of the wrong sign.

This is clearly unacceptable. Technically, one
can solve the problem by adding a constant $+\mu^4/4 \lambda$ to $V$, so
that $V$ is equal to zero at the minimum ({i.e.,} $\Lambda_{
bare} = 2 \pi G_N \mu^4/\lambda)$.  However, with our current
understanding there is no reason for $\Lambda_{bare}$ and
$\Lambda_{SSB}$ to be related. The need to invoke such an
incredibly fine-tuned cancellation to 50 decimal places is probably the most
unsatisfactory feature of the standard model.
The problem becomes even worse in superstring theories,
 where one expects a vacuum energy
of ${\cal O} (M_P^4)$ for a generic point in the landscape, leading to $\Lambda_{obs} \gtrsim 10^{123} |\Lambda_{obs}|$.
The situation is almost as bad in grand unified theories.

So far no compelling solution to the cosmological constant problem has emerged. One intriguing
possibility invokes the anthropic (environmental) principle~\cite{Barrow:109141,Rees:2004av,Hogan:1999wh}, i.e., that a much larger or smaller value 
of $|\Lambda_{cosm}|$
would not have allowed the possibility  for life to have evolved because the Universe would have expanded or 
recollapsed too rapidly~\cite{Weinberg:1988cp}.
 This would be a rather meaningless argument unless (a) Nature somehow allows a large variety
 of possibilities for $|\Lambda_{cosm}|$ (and possibly other parameters or principles)
 such as in different vacua, and (b) there is some mechanism to try all or many of them. In recent years it has been suggested that both of these needs may be met.
There appear to be an enormous landscape of possible superstring vacua~\cite{Bousso:2000xa,Kachru:2003aw,Susskind:2003kw,Denef:2004ze}, with no obvious physical principle to choose one over the other. Something like eternal inflation~\cite{Linde:1986fc} could provide the means to sample them, so that only the environmentally suitable vacua lead to long-lived Universes suitable for life.
These ideas are highly controversial and are currently being heatedly debated.

\subsection*{The New Ingredients}
It is now clear that the standard model  requires a number of new ingredients.
These include
  \begin{itemize}
\item {\bf A mechanism for small neutrino masses.} The most popular possibility
is the minimal seesaw model, implying Majorana masses, but there are other
plausible mechanisms for  either small Dirac or Majorana masses~\cite{GonzalezGarcia:2002dz,Langacker:2005pfa,Mohapatra:2005wg,GonzalezGarcia:2007ib}.
\item {\bf A mechanism for the baryon asymmetry.}
The standard model has neither the nonequilibrium condition nor sufficient $CP$
violation to explain the observed asymmetry between baryons and antibaryons in the Universe~\cite{Sakharov:1967dj,Bernreuther:2002uj,Dine:2003ax}\footnote{The third necessary ingredient, baryon number nonconservation, is present in the SM because of
non-perturbative vacuum tunnelling (instanton) effects~\cite{'tHooft:1976up}. These are negligible at zero temperature
where they are exponentially suppressed, but
important at high temperatures due to thermal fluctuations (sphaleron configurations), before or during the electroweak
phase transition~\cite{Klinkhamer:1984di,Kuzmin:1985mm}.}.
One possibility involves the out of equilibrium decays of superheavy Majorana right-handed neutrinos
({leptogenesis}~\cite{Fukugita:1986hr,Davidson:2008bu}),
as expected in the minimal seesaw model. Another involves a strongly first order electroweak phase transition ({electroweak baryogenesis}~\cite{Trodden:1998ym}).
This is not expected in the standard model, but could possibly be associated with loop effects in the minimal
supersymmetric extension (MSSM) if one of the scalar top quarks is sufficiently light~\cite{Carena:2008rt}. However, it is most likely in extensions of the MSSM involving 
SM singlet Higgs fields that can generate a dynamical $\mu$ term, which can easily lead to 
strong first order transitions at tree-level~\cite{Barger:2007ay}. Such extensions would likely yield signatures observable at the LHC.
Both the seesaw models and the singlet extensions of the MSSM
 could also provide the needed new sources of $CP$ violation. Other possibilities for the baryon asymmetry include the
 decay of a coherent scalar field, such as a scalar quark or lepton in supersymmetry (the {Affleck-Dine mechanism}~\cite{Affleck:1984fy}), or $CPT$  violation~\cite{Cohen:1987vi,Davoudiasl:2004gf}.
 Finally, one cannot totally dismiss the possibility that the asymmetry is simply due to an initial condition on the big bang. However, this possibility disappears if the universe underwent a period of rapid
 {inflation}~\cite{Lyth:1998xn}.
 \item {\bf What is the dark energy?}
 In recent years a remarkable concordance of cosmological observations involving the 
 cosmic microwave background radiation (CMB), acceleration of the Universe as determined by
 Type Ia supernova observations, large scale distribution of galaxies and clusters, and big bang nucleosynthesis
 has allowed precise determinations of the cosmological parameters~\cite{Kolb:206230,Peebles:250028,Dunkley:2008ie,Amsler:2008zz}: the Universe is close to flat, with some form of {dark energy}
 making up about 74\% of the energy density. {Dark matter} constitutes $\gtrsim 21$\%, while ordinary
 matter (mainly baryons) represents only about 4-5\%.  
 The mysterious dark energy~\cite{Peebles:2002gy,Frieman:2008sn,Martin:2008qp}, which is the most important contribution to the energy density and leads to the  acceleration of the expansion of the Universe,
 is  not accounted for in the SM.
  It could be due to a
cosmological constant that is incredibly  tiny on the particle physics scale, or  to a slowly time varying field ({quintessence}). Is the acceleration somehow related to an
earlier and much more dramatic period of {inflation}~\cite{Lyth:1998xn}? If it is associated with a time-varying field, could it be connected
with a possible time variation of coupling ``constants''~\cite{Uzan:2002vq}?
\item {\bf What is the dark matter?}
 Similarly, the standard model has no explanation for the observed dark matter, which contributes much more
 to the matter in the Universe than the stuff we are made of. It is likely, though not certain, that the dark matter
 is associated with elementary particles. An attractive possibility is {weakly interacting massive particles} (WIMPs),
 which are typically particles in the $10^2-10^3$ GeV range with weak interaction strength couplings, and which lead naturally
 to the observed matter density. These could be associated with the lightest supersymmetric partner (usually a neutralino)
 in supersymmetric models with $R$-parity conservation, or analogous stable particles in Little Higgs or universal
 extra dimension models. There are a wide variety of variations on these themes, e.g., involving very light
 gravitinos or other supersymmetric particles. There are many searches for WIMPs going on, including direct searches for the recoil produced by scattering of  Solar System WIMPs, indirect searches for WIMP annihilation products, and searches for WIMPs produced at accelerators~\cite{Jungman:1995df,Bertone:2004pz,Hooper:2008sn}. Axions, perhaps associated
 with the strong $CP$ problem or with string vacua~\cite{Svrcek:2006yi}, are another possibility. Searches for axions produced in the Sun, in the laboratory, or from the early universe are currently underway~\cite{Asztalos:2006kz,Kim:2008hd}.
  \item {\bf The suppression  of flavor changing neutral currents, proton decay, and
 electric dipole moments.}
 The standard model has a number of accidental symmetries and features
 which forbid proton decay, preserve lepton number and lepton family number (at least for vanishing neutrino
 masses), suppress transitions such as $K^+\ra \pi^+ \nu \bar \nu $ at tree-level,
 and lead to highly suppressed electric dipole moments for the $e^-$, $n$, atoms, etc. 
 However, most extensions of the
 SM have new interactions which violate such symmetries, leading to potentially serious problems with
 FCNC and EDMs. There seems to be a real conflict between attempts to deal with the Higgs/hierarchy
 problem and the prevention of such effects.
 
 Recently, there has been much discussion of {minimal flavor violation}, which is the hypothesis that all flavor
 violation, even that which is associated with new physics, is proportional to the standard model Yukawa matrices~\cite{D'Ambrosio:2002ex,Nir:2007xn},
 leading to a significant suppression of flavor changing effects.
\end{itemize}

\section*{Acknowledgements}
I am grateful to Tao Han 
for inviting me to give these lectures.
This work was supported by the organizers of TASI2008, the IBM Einstein Fellowship, and by NSF
grant PHY-0503584.

\bibliographystyle{ws-procs9x6}
\bibliography{bib_pgl}
\end{document}